\documentclass[useAMS,usenatbib]{mn2e}
\usepackage{graphicx}
\usepackage{epstopdf}
\usepackage{amssymb}
\usepackage{color}
\newcommand{\rxte}{{\it RXTE}}
\newcommand{\rxtefan}{{\it Rossi X-ray Timing Explorer}}
\newcommand{\rxteasm}{{\it RXTE}/ASM}
\newcommand{\rxtepca}{{\it RXTE}/PCA}
\newcommand{\rxtehexte}{{\it RXTE}/HEXTE}
\newcommand{\pca}{{PCA}}
\newcommand{\pcafan}{{Proportional Counter Array (PCA)}}
\newcommand{\asm}{{ASM}}
\newcommand{\asmfan}{{All-Sky Monitor (ASM)}}
\newcommand{\hexte}{{HEXTE}}
\newcommand{\hextefan}{{High Energy X-ray Timing Experiment (HEXTE)}}
\newcommand{\integral}{{\it INTEGRAL}}

\newcommand{\swiftxrt}{{\it Swift}/XRT}

\newcommand{\compps}{{\tt compps}}

\newcommand{\diskbb}{{\tt diskbb}}
\newcommand{\isis}{{\tt ISIS}}

\newcommand{\cgrobatse}{{\it CGRO}/BATSE}
\newcommand{\batsefan}{{Burst And Transient Source Experiment (BATSE)}}
\newcommand{\cgroosse}{{\it CGRO}/OSSE}

\newcommand{\cgrofan}{{\it Compton Gamma Ray Observatory (CGRO)}}
\newcommand{\clc}{{\tt correctlc\_v0.1}}

\newcommand{\ftools}{{\tt FTOOLS\/ v6.5}}

\title[The HID of Cyg X-3]{The Hardness-Intensity Diagram of Cygnus X-3: Revisiting the Radio/X-Ray States}

\author[K. I. I. Koljonen et al.]
{K.~I.~I.~Koljonen$^{1}$\thanks{email: karri@kurp.hut.fi}, D.~C.~Hannikainen$^{1,2}$, M.~L.~McCollough$^{3}$, G.~G.~Pooley$^{4}$, 
\and S.~A.~Trushkin$^{5}$
\\
$^{1}$Aalto University Mets\"ahovi Radio Observatory, Mets\"ahovintie 114, 02540 Kylm\"al\"a, Finland \\
$^{2}$Tuorla Observatory, University of Turku, V\"ais\"al\"antie 20, 21500 Piikki\"o, Finland \\
$^{3}$Smithsonian Astrophysical Observatory, 60 Garden Street, Cambridge, MA 02138-1516, USA \\ 
$^{4}$Astrophysics Group, Cavendish Laboratory, 19 J. J. Thomson Avenue, Cambridge CB3 0HE \\
$^{5}$Special Astrophysical Observarory RAS, Karachaevo-Cherkassian res, Nizhnij Arkhyz, 36916, Russia }

\date{Accepted 2010 March 16.  Received 2010 March 16; in original form 2009 December 23}

\begin{document}

\pagerange{\pageref{firstpage}--\pageref{lastpage}}
\pubyear{2010}

\maketitle

\label{firstpage}

\begin{abstract} 

Cygnus X-3 is one of the brightest X-ray and radio sources in the Galaxy, and is well known for its erratic behaviour in X-rays as well as in the radio, occasionally producing major radio flares associated with relativistic ejections. However, even after many years of observations in various wavelength bands Cyg X-3 still eludes clear physical understanding. Studying different emission bands simultaneously in microquasars has proved to be a fruitful approach towards understanding these systems, especially by shedding light on the accretion disc/jet connection. We continue this legacy by constructing a hardness-intensity diagram (HID) from archival \rxtefan\/ data and linking simultaneous radio observations to it. We find that surprisingly Cyg X-3 sketches a similar shape in the HID to that seen in other transient black hole X-ray binaries during outburst but with distinct differences. Together with the results of this analysis and previous studies of Cyg X-3 we conclude that the X-ray states can be assigned to six distinct states. This categorization relies heavily on the simultaneous radio observations and we identify one new X-ray state, the hypersoft state, similar to the ultrasoft state, which is associated to the quenched radio state during which there is no or very faint radio emission. Recent observations of GeV flux observed from Cyg X-3 \citep{tavani,corbel} during a soft X-ray and/or radio quenched state at the onset of a major radio flare hint that a very energetic process is at work during this time, which is also when the hypersoft X-ray state is observed. In addition, Cyg X-3 shows flaring with a wide range of hardness.

\end{abstract}


\section{Introduction}

Cygnus X-3 is a well-known X-ray binary (XRB) located in the plane of the Galaxy. Its discovery dates back to 1966 \citep{giacconi} but the nature of the system has remained a mystery despite extensive multiwavelength observations throughout the years. Its X-ray \citep{parsignault} show a very strong and infrared \citep{mason} weaker 4.8-hour orbital modulation, typical of low-mass XRBs. However, infrared observations indicate that its mass-donating companion is a Wolf-Rayet (WR) star \citep{keerkwijk}, which would make it a high-mass XRB. Also unlike most other XRBs, Cyg X-3 is relatively bright in the radio virtually all of the time and it undergoes giant radio outbursts with strong evidence of jet-like structures moving away at relativistic speeds \citep{molnar,schalinski,mio}.

The distance to Cyg X-3 is rather poorly constrained due to heavy absorption in the Galactic plane totally obscuring the optical wavelengths. Depending on the method, the distance to Cyg X-3 is reported to be $>9.2$ kpc \citep{dickey}, 8 kpc \citep{predehl1} and $9^{+4}_{-2}$ \citep{predehl2}. Moreover, it is quite possible that the accretion disc is enshrouded partially or wholly in the strong stellar wind of the WR companion (\citealt{fender1}; \citealt{szostek2}; \citealt{vilhu2}). 

The nature of the compact object is not certain, but it is thought to be a black hole due to its spectral resemblance to other black hole XRB systems, such as GRS 1915$+$105 and XTE J1550$-$564 (see e.g. \citealt{szostek3}, hereafter S08; \citealt{hjalmarsdotter2}). Also there is no evidence of a neutron star system producing such massive radio outbursts (up to 20 Jy, \citealt{waltman2}) as observed from Cyg X-3. Cyg X-3 is a unique system in our Galaxy, but recent observations of two XRBs in IC 10 \citep{prestwich} and NGC 300 \citep{carpano}, both containing a WR companion, show strong evidence of a black hole primary. It is worth noting that these systems may represent a crucial link towards the evolution of a double black hole binary.

The X-ray spectra of Cyg X-3 are notoriously complex. Basically, Cyg X-3 exhibits the canonical X-ray states seen in other XRBs, namely the high/soft (HS) and low/hard (LH) states, in addition to the intermediate, very high and ultrasoft states (e.g. \citealt{szostek1}, hereafter S04; \citealt{hjalmarsdotter2}). However, in Cyg X-3, the strong radio emission is also classified into states of its own (\citealt{waltman3,mccollough1}, hereafter M99). Recently these X-ray and radio states were implemented into a more unified picture as presented in S08: the radio/X-ray states. The X-ray emission has been found to be linked to radio emission in Cyg X-3. M99 found that during periods of flaring activity in the radio the hard X-ray (HXR) flux switches from an anti-correlation to a correlation with the radio. In addition, the HXR flux has been shown to anti-correlate with the soft X-rays (SXR) in both canonical X-ray states \citep{mccollough}.

In February 1997 a large radio flare ($\sim$ 10 Jy) was observed by the Green Bank Interferometer (GBI) and the Ryle Telescope after a period of quenched emission. At the same time, the \batsefan\/ onboard the \cgrofan\/ detected a flare in the HXR that showed a strong correlation with the radio. The flare in the HXR was preceded by several days of very low HXR flux (below the \cgrobatse\/ one-day detection limit). The flare triggered a Very Long Baseline Array (VLBA) observation to obtain high resolution radio images of Cyg X-3 during the major flare. The resulting VLBA observations show an expanding one-sided jet \citep{mio} that was found to have a velocity of $\sim$ 0.81$c$ and an inclination of $\sim$ 14 degrees to our line of sight.

An important tool in understanding the nature of the transient black hole systems is the hardness-intensity diagram (HID, e.g. \citealt{fender2} and references therein). Thus it should be natural to look at Cyg X-3's HID and compare it to other black hole systems, all the while bearing in mind that Cyg X-3 does not behave like a transient black hole XRB in outburst. Nevertheless, we find the HID to be a useful tool even in this case. \citet{smit} have taken a first look at Cyg X-3's behaviour in a HID using EXOSAT data, where they identified two distinct branches with a possible third branch connecting these two. In this paper we further classify the different radio/X-ray states by constructing a HID for Cyg X-3 from the X-ray data and adding the radio dimension in order to shed more light on the disc/jet connection in the system. This is made possible by the large amount of data available from the \rxtefan\/ (\textit{RXTE}) archive (totaling to $\sim$ Msec) and simultaneous radio observations from the GBI, RATAN-600 and Ryle telescopes (their monitoring programs of microquasars). 

We describe the X-ray and radio data in Section 2. In Section 3 we summarize the disc-jet connection of Cyg X-3 and the HID of black hole systems. In Section 4 we summarize and refine the radio/X-ray states, in addition to presenting the HID including radio observations. We discuss the ramifications of our results in Section 5 and summarize the paper in Section 6. 

\section{Observations}

\subsection{Pointed Observations}

We have analyzed, following standard procedures using \ftools\/, 135 archive \rxte\/ observations\footnote{Observations from P10126 (8 obs.), P20099 (6 obs.), P20101 (10 obs.), P30082 (6 obs.), P40061 (17 obs.), P40422 (1 obs.), P50062 (22 obs.), P70062 (11 obs.), P91090 (30 obs.), P91412 (1 obs.), P93434 (5 obs.) and P94328 (18 obs.)} for the hardness-intensity diagram. The observations are listed in Table \ref{list1} together with exposure times of the \pcafan\/ and \hextefan\/ as well as corresponding radio/X-ray states as defined in Section 4.2 and radio flux during or near the observation in question (maximum of half a day offset). The \pca\/ lightcurves were extracted from channels 3--11 and 24--37 corresponding to energy ranges 3--6 keV and 10--15 keV. These bands were chosen so as to probe different emission regions yet at the same time yielding a maximum number of counts; in this case the SXR range represents the disc and the HXR range the Comptonized part of the spectrum. Each lightcurve was also corrected with \clc\/ to scale it to a single, mean PCU, so that the lightcurves are comparable to one another. Then each lightcurve was binned into a 0.01 day (864 sec) bin. The spectra of each pointing were extracted using the standard method as described in the \rxte\/ Cook Book. For the \pca\/ data, bins below 3 keV and above 20 keV were ignored, as were bins in the \hexte\/ data above 200 keV. In addition, the \pca\/ data were grouped to a minimum of 30 counts/bin and the \hexte\/ data were grouped to a minimum of 2/3/5 sigma significance per bin depending on the radio/X-ray state of the system: 2 for the hypersoft (SXR spectra), 3 for the flaring and transition spectra (intermediate X-ray spectra) and 5 for the quiescent spectra (HXR spectra). A summed spectrum was also created for each radio/X-ray state (see Section 4) and grouped to a minimum of three (hypersoft) or five sigma (the rest). See Section 4 for the description of different states. The spectral fit of the hypersoft state was done using \isis\/ (Interactive Spectral Interpretation System, \citealt{houck}) and a 0.5\% systematic error was added to the \pca\/ data. 

\subsection{Monitoring Observations}

In this work we have also used monitoring data from the \asmfan\/ onboard \rxte\/ and from the following radio observatories: Ryle (15 GHz), GBI (2.25 GHz and 8.3 GHz, \citealt{waltman1,waltman2,waltman3}) and the RATAN-600 (11.2 GHz, \citealt{thruskin}). For the \asm\/, we used the dwell lightcurves in three different modes: single dwell, phase-selected and phase-selected daily-binned dwells. For the phase-selected dwells, we calculated the phase of each dwell and selected only those with phase inside 0.4--0.6. This is done because the orbital modulation alters the lightcurve by order of two or three but inside the chosen phase range the SXR flux is rather constant \citep{vilhu1}. We use a cubic ephemeris \citep{singh} for the phase determination. Secondly, we bin these phase-selected dwells into a one-day bin of weighted average. We use the differently binned dwells to probe the variability of SXR for different radio/X-ray states (see Fig. \ref{asmhid}).

\section{Disc-jet Connection}

The disc-jet connection is one of the most intriguing aspects of accretion physics and it is not yet well established what is the underlying mechanism for launching the energetic and tightly collimated jets emanating from the central regions of the XRB systems. Microquasars offer us an apt laboratory for studying this connection and Cyg X-3 is a prime example of a system capable of launching relativistic jets observed as major flares in the radio. One established method for probing the disc-jet connection in transient XRBs during outburst is the HID (e.g. \citealt{fender2} and references therein) linked to simultaneous radio observations. 

In this paper we create a HID for Cyg X-3. Recently, there have been studies of the radio/X-ray states of Cyg X-3 (S08) and we associate these states as well as the X-ray spectral states presented in S04 to the HID in this paper to form a unified picture of the system. For reference we have gathered the different classification methods of the radio and X-ray spectral states in Table \ref{cygstates} along with the new ones proposed in this paper. We strongly urge the reader to follow this table along with the text since the spectral states, and the classifications, are highly complex and may appear otherwise confusing. Based on daily examination of more than a decade of Ryle and \rxteasm\/ data Cyg X-3 spends roughly half the time in radio/X-ray quiescence and the rest in a flaring state. Major radio flaring periods are of prime importance for studying the relationship between the accretion disc and the radio jets. We will also explore the evolution of the X-ray spectra during the major radio flares together with soft X-ray, hard X-ray and radio evolution of the monitoring data. We start by examining the historic studies of the various radio/X-ray connections.    

\subsection{Radio, Soft X-ray, and Hard X-ray Relationships in Cygnus X-3}

In the first study of its kind dedicated to classifying Cyg X-3's behaviour in the X-ray and radio, \citet{mccollough} compared simultaneous HXR (\cgrobatse\/ 20--100 keV) observations with radio (GBI 2.25 and 8.3 GHz, Ryle 15 GHz) and SXR (\rxteasm\/ 1-12 keV) observations. This work was further continued in M99. The following discoveries hold over all of the overlaps between the data sets: 

(a) During times of moderate radio brightness ($\sim$ 100 mJy), and low variability in the radio and X-ray fluxes, the HXR flux anti-correlates with the radio. It is during this time that the HXR reaches its highest level. 

(b) During periods of flaring activity in the radio the HXR flux switches from an anti-correlation to a correlation with the radio. In particular, for major radio flares and the quenched radio emission (very low radio fluxes of 10--20 mJy) which precedes them the correlation is strong. 

(c) The HXR flux has been shown to anti-correlate with the SXR. This occurs in both the low/hard and high/soft X-ray states. 

(d) The occurrences of flaring periods in the radio during the high/soft X-ray states \citep{watanabe} were confirmed. 

(e) It has been found from the \cgrobatse\/ and \cgroosse\/ data that the spectrum of Cyg X-3 (in the 20--100 keV band) hardens during times of radio flaring \citep{mccollough3}. 

\subsection{Hardness-Intensity Diagrams of Black Hole Systems}

A typical black hole HID representing the transient outbursting cycle has a Q-type shape (\citealt{fender2}, their Fig. 7) that can be divided into four main areas or states: quiescent, low/hard (LH), very high/intermediate (VHS/IS) and high/soft (HS) state. These areas/states are briefly reviewed in the following: 

\textit{Quiescent state (lower right in the diagram)}: The source flux is very low in both the X-ray and radio with the spectrum described by a hard power law; 

\textit{LH (right side)}: The source can be bright in the X-rays and its spectrum is described by a power law with exponential cut-off at $\sim$ 100 keV. The radio emission has a flat spectrum and appears as a compact jet; 

\textit{VHS/IS (top)}: During this state there is a prominent thermal disc component and a steep power law without a cut-off; 

\textit{HS (left side)}: This state is thermally dominated and typically there is no power law component, or if present, it is very weak. There is also a line that demarcates the VHS/IS region from the HS region that is referred to as the \textit{jet line}. When the source crosses this line going from right to left there is a sudden cut-off of radio emission. It is during this transition that strong radio flares occur. 

\section{The Hardness-Intensity Diagram of Cygnus X-3}

\begin{figure*}
\begin{center}
\includegraphics[width=0.98\textwidth]{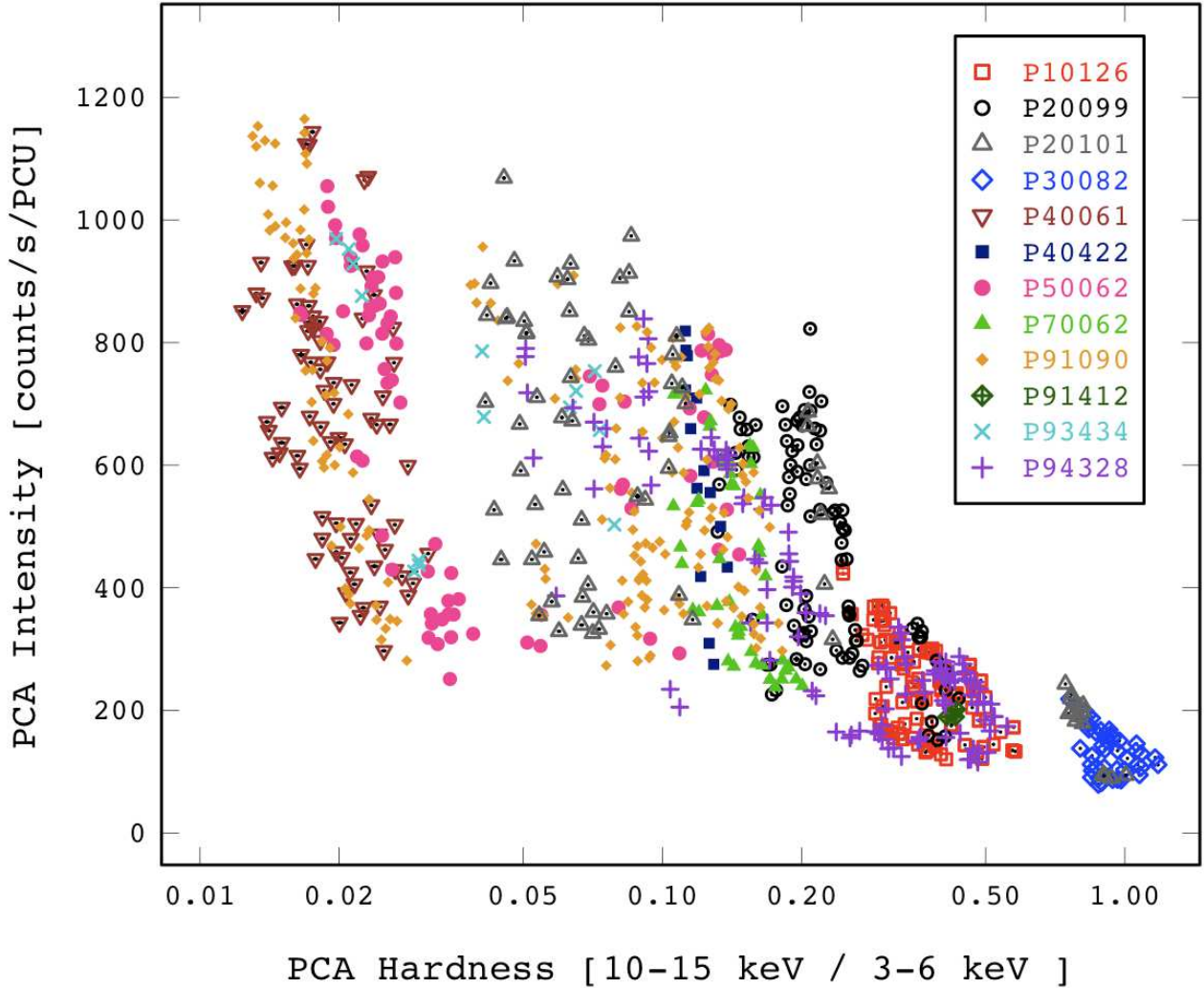}
\caption{A hardness-intensity diagram (HID) for Cyg X-3 from 135 pointed \pca\/ observations. Different observation periods are plotted with different colour and symbol according to the legend. Note the similarity in appearance to the HID for black hole transient systems \citep{fender2}. See the text for the noted differences.} \label{hid}
\end{center}
\end{figure*}    

Fig. \ref{hid} shows the HID for Cyg X-3. The bands were chosen so as to probe different emission regions, with the SXR (3--6 keV) range representing the disc and HXR (10--15 keV) range representing the Comptonized part of the spectrum. This division is also similar to work done on other black hole XRBs, e.g. 3.8--6.3 keV and 6.3--10.5 keV in \citet{fender2} (except for GRS 1915+105) or more recently 3--6 keV and 6--10 keV in \citet{dunn}. The harder band was chosen in this study to ensure that the disc (1.3--1.6 keV in the HS state) does not significantly contribute to it.

At first glance the plot appears surprisingly similar to other black hole XRB HIDs, but there are some very important differences. First of all, Cyg X-3 does not show hysteresis in the HID \citep{hjalmarsdotter2} but simply increases in intensity as the spectrum softens (maximum \rxtepca\/ intensity 1200 counts/s/PCU or $\sim1\times 10^{-8} \, \rm{erg}/\rm{cm}^{2}/\rm{s}$ in the 3--15 keV band, in the flaring/hypersoft states as compared to 200 counts/s/PCU or $\sim 2 \times 10^{-9} \, \rm{erg}/\rm{cm}^{2}/\rm{s}$ in the quiescent state). In black hole XRBs the LH state is present throughout the right branch, whereas in Cyg X-3 the LH state is confined to the foot of the Q. The right branch in this case is dominated by X-ray flaring in Cyg X-3. In addition, the flaring data are also spread into different hardnesses which wholly fill the inside area of the Q.  

\subsection{Variability in Cygnus X-3's HID}

\begin{figure}
\begin{center}
\includegraphics[width=0.5\textwidth]{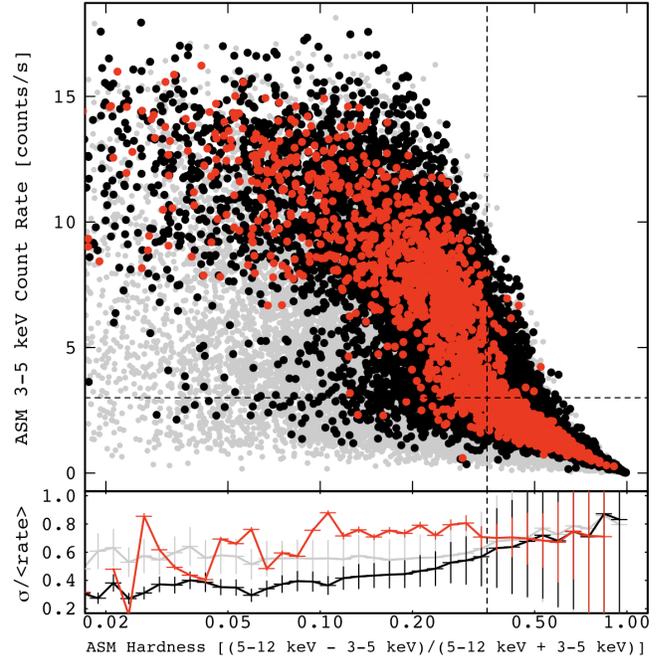}
\caption{\textit{Top:} a hardness-intensity diagram (HID) for Cyg X-3 from all available \asm\/ data. All \asm\/ dwells are plotted as grey points, black points represent the phase-selected (0.4--0.6) dwells and red points the phase-selected daily binned dwells. Two dashed lines have been added: a horizontal line representing the transitional value defined in S08 demarcating the HID into quiescent states (below the line) and flaring states (above the line) based on the \asm\/ 3--5 keV flux, and a vertical line similarly demarcating the HID into quiescent states (right side of the line) and flaring states (left side of the line) based on the variability patterns of both the radio and SXR (S08). \textit{Bottom:} the SXR variability index for different hardnesses. Colours are as defined in the top panel.} \label{asmhid}
\end{center}
\end{figure}

As we see in the Cyg X-3 HID, there is noticeable scatter present in the intensity for each value of hardness and the most striking transition happens when the system moves in or out of the flaring region. A vertical line in Fig. \ref{asmhid} (also marked in Figs. \ref{hidradio} and \ref{hidroutes}) marks the change in the system from the LH state to the HS state or vice versa, and we see a corresponding increase/decrease in SXR flux (see also spectra of different states of Cyg X-3 in Fig. \ref{plotall} and note the difference in the SXR component between the transition state and the FHXR state). The scatter in the intensity is partly due to the strong orbital motion detected in the system (e.g. \citealt{vilhu1}) which affects the intensity approximately by a factor of two. The orbital modulation impacts also on the hardness ratios as noted by \citet{smit}. On top of that we expect X-ray flaring occurring in the system which will increase the variability in the intensity by some amount. 

In order to study the X-ray variability in more detail we constructed a HID for Cyg X-3 from all available \asm\/ data, selecting the 3--5 keV and 5--12 keV bands so as to obtain similar hardness ratios as for the pointed observations (Fig. \ref{asmhid}). In the top panel of Figure \ref{asmhid} all \asm\/ dwells are plotted as gray points, phase-selected (0.4--0.6) dwells as black points and phase-selected daily-binned dwells as red points. We note that the HID is of the same shape as obtained from the pointed observations and the HID appears to be more continuum-like rather than displaying distinct states. Also the effect of orbital modulation (compare grey points with black and red) is notable and constitutes about half of the scatter in intensity. The transition value of $\sim$ 3 counts/s or $\sim1\times10^{-9} \, \rm{erg}/\rm{cm}^{2}/\rm{s}$ in the \rxteasm\/ 3--5 keV band as defined in S08 is marked as a horizontal, dashed line, demarcating the HID into two regions: quiescent states are below the line and flaring states above the line as defined in S08. A similar demarcating line can be inserted also vertically based on the variability patterns of both the radio and SXR in a quiescent state (S08), corresponding approximately to where the \asm\/ 3--5 keV count rate starts increasing more rapidly (\asm\/ hardness $\sim$ 0.35), the quiescent states now on the right side of this line and the flaring states on the left. In the bottom panel the variability index is plotted as a function of the hardness. The index is defined as the ratio between the standard deviation and mean rate of the selected dwells (summed \asm\/ rate 1--12 keV), in this case from the same hardness band. The index increases with the hardness except in the case of the daily binned dwells, where it stays roughly constant. However, the SXR variability in the quiescent states is not well constrained due to the low number of counts. 

\subsection{Revised Classifications of Cygnus X-3's Radio/X-ray spectral States}

Using all the \rxte\/ data available from 1996--2000, S04 divided the Cyg X-3 spectra into five groups numbered 1--5. Groups 1--2 were classified as representing the hard state, groups 3--4 the intermediate states, and group 5 the soft state. The criterion used for this classification was the X-ray flux at 20 keV. In a more recent study, S08 added radio data to the study of S04, and thus refined the definition of the X-ray states taking into account the radio-activity. They employed a transitional flux level to distinguish between the states, which was $\sim$ 3 counts/s or $\sim1\times10^{-9} \, \rm{erg}/\rm{cm}^{2}/\rm{s}$ in the \rxteasm\/ 3--5 keV band and $\sim$ 0.25 Jy in the 8.5 GHz GBI frequency band. In the following, we present our classification of the radio/X-ray states of Cyg X-3, closely linked to the studies of S04 and S08, but with a new take on the classical HID and including additional radio data. We will describe each one of our X-ray states in relation to those presented in the studies discussed above.

\begin{figure*}
\begin{center}
\includegraphics[width=0.95\textwidth]{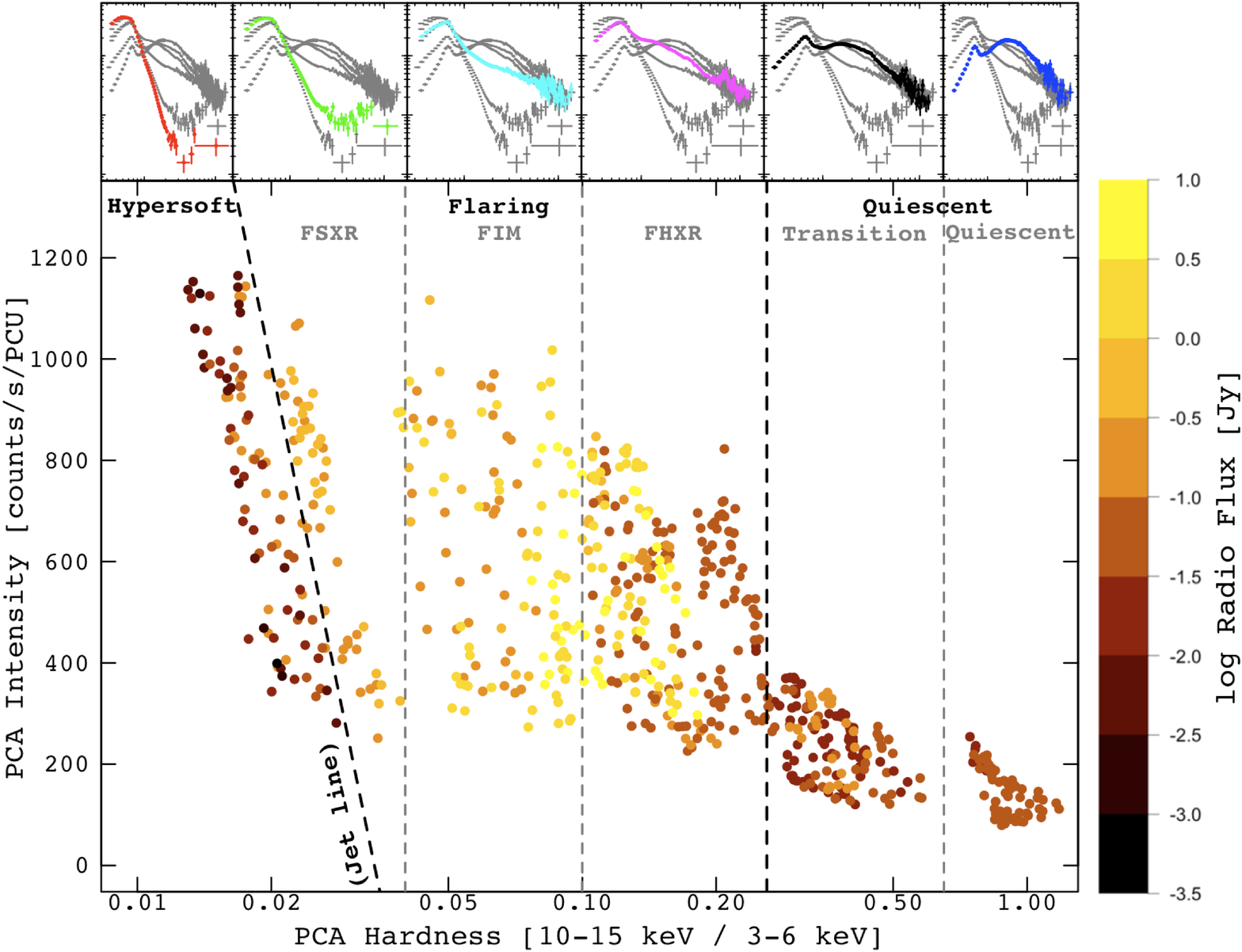}
\caption{The same hardness-intensity diagram as in Fig. \ref{hid}, but now the data points which have simultaneous or nearly simultaneous radio observations have been coloured according to the radio flux, darker colour referring to lower flux. This shows clearly the different radio/X-ray states in the HID. Also, the plot is divided into six different areas which correspond to six different X-ray states of Cyg X-3. These state spectra are plotted in the upper part of the figure each on top of its own area. The spectra are coloured as in Fig. \ref{plotall}. The division between the hypersoft state and the FSXR state is labelled as the jet line. When Cyg X-3 crosses this line from the hypersoft state it produces a major flare. However, no flares have been observed when the systems crosses the line in the other direction.} \label{hidradio}
\end{center}
\end{figure*}

In Fig. \ref{hidradio}, the HID for Cyg X-3 is plotted using a new approach: the data points with (near-)simultaneous radio coverage have been coloured according to the strength of the radio flux density (data from Table \ref{list1}), with darker points representing a weaker radio source and lighter points a brighter radio source. Thus the radio/X-ray states are clearly revealed in the HID and we divide them according to the X-ray hardness and/or radio flux into three distinct areas that we call hypersoft, flaring and quiescent. In addition, the flaring state can be further subdivided according to X-ray hardness and the shape of the spectra: flaring/soft X-ray (FSXR), flaring/intermediate (FIM) and flaring/hard X-ray (FHXR).  The quiescent state can also be subdivided into two regions: quiescent and transition. These subdivisions are rather ad hoc since the flaring spectra and the quiescent spectra are more or less continuous as is seen in Fig. \ref{asmhid}, but the division is made so that we can compare the spectra with those presented in S04 and S08. 

The spectra for all states are shown in Fig. \ref{plotall} (average of each state, all states plotted in the same figure), and reproduced above the HID in Fig. \ref{hidradio}. The descriptions of the states from this study and those of S04 and S08 with as close correspondence as possible are tabulated in Table \ref{cygstates}. In the following we describe the states as defined in this paper, in conjunction with the previous studies mentioned above (S04, S08).

\begin{table*}
\caption{Different classification methods of Cyg X-3 X-ray spectra with corresponding radio activity.} \label{cygstates}
\begin{tabular}{cccccc}
\hline
\hline
{\rm Canonical X-ray states} & {\rm Radio states}$^{a}$ & \multicolumn{2}{c}{{\rm X-ray states of S04}$^{b}$} & {\rm radio/X-ray states of S08}$^{c}$ & This paper \\ 
\hline 
& & group: & name: \\
\\
{\rm low/hard} & {\rm quiescent} & \raisebox{1.5ex}{\rm 1} & \raisebox{1.5ex}{\rm hard} & \raisebox{1.5ex}{\rm quiescent} & \raisebox{1.5ex}{\rm quiescent} \\
& & \raisebox{1.5ex}{\rm 2} & \raisebox{1.5ex}{\rm intermediate} & \raisebox{1.5ex}{\rm minor flaring} & \raisebox{1.5ex}{\rm transition} \\
\\
{\rm intermediate} & {\rm minor flaring} & 3 & {\rm very high} & \raisebox{2.0ex}{\rm suppressed} & {\rm FHXR} \\
& & & & \raisebox{1.5ex}{\rm ``post-flare"} \\
\\
& {\rm major flaring} & \raisebox{1.5ex}{\rm 4} & \raisebox{1.5ex}{\rm soft non-thermal} & \raisebox{1.5ex}{\rm major flaring} & \raisebox{1.5ex}{\rm FIM} \\
{\rm high/soft} & & {\rm 5} & {\rm ultrasoft} & {\rm quenched} & {\rm FSXR} \\
& {\rm quenched} & & & & {\rm hypersoft}\\
\hline
\end{tabular}
\begin{list}{}{}
\item[$^{\mathrm{a}}$] \citet{waltman3} and \citet{mccollough1}.
\item[$^{\mathrm{b}}$] \citet{szostek1} and \citet{hjalmarsdotter2}.
\item[$^{\mathrm{c}}$] Szostek, Zdziarski \& McCollough (2008).
\end{list}
\end{table*}

\subsubsection{Quiescent State}

The quiescent region comprises the lower right corner of the HID, bounded by the flaring region on the left side. Based on the Ryle and \rxteasm\/ 3-5 keV data and imposing a cut-off of 200 mJy in the radio and 3 counts/s in the ASM data to demarcate the quiescent from the flaring states, Cyg X-3 is found to spend $\sim$ 52 \% of the time in the quiescent states and the rest in the flaring states. The quiescent region corresponds to the quiescent state of S08 and the spectra of this region correspond to the spectra of groups 1 and 2 of S04 and therefore are divided here into two states: the quiescent state and the transition state.

The quiescent state has a low number of counts in the SXR and therefore the level of SXR variability is not well constrained (see Fig. \ref{asmhid}). The HXR and radio variability are low compared to the flaring states. The radio spectrum is optically thick and the radio varies in a correlated manner with the SXR below the transition level as defined in S08. In addition, the HXR flux anti-correlates with the radio (M99). In the quiescent region Cyg X-3 is still relatively bright in both the X-ray and the radio (60--200 mJy) as compared to other black hole XRB systems in this state. A recent study \citep{hjalmarsdotter1} of the low/hard state of Cyg X-3 using \integral\/ found: (1) the HXR spectrum has a cut-off at $\sim$ 20 keV (for most black hole XRBs in a state similar to this one the cut-off occurs at $\sim$ 100 keV), this cut-off at lower-than-expected energies has recently been explained as being due to Compton downscattering from a nearby medium \citep{zdziarski}; (2) there is a significant contribution from non-thermal Comptonization, which is usually observed only in soft states; and (3) the luminosity of the hard state is significantly higher than other XRBs in this state, implying a model much more radiatively efficient than the standard ADAF models or a mass of the compact object $\gtrsim$ 20M$_{\odot}$. 

\subsubsection{Transition State}

The transition state is between the quiescent and flaring regions in the HID. In the transition region the radio flux starts to increase and the power law component of the spectrum starts to flatten shifting the HXR spectral peak towards softer energies, i.e. the X-ray hardness decreases and the source moves to the left in the HID. However, the overall X-ray spectrum is still similar to the quiescent/LH state spectrum and the variability in intensity remains similar to the quiescent state. Also the HXR variability at the level of the quiescent state and the movement in the \asm\/ HID (Fig. \ref{asmhid}) appears to be continuous. However, the radio variability increases to the level of flaring states indicating the commencement of minor radio flaring. 

\subsubsection{Flaring/Hard X-ray State (FHXR)}

The flaring region comprises the main part of the Cyg X-3 HID. It is bound by the transition region on the right side and the hypersoft region on the left side. The most striking aspect of the flaring states is the increased radio and X-ray emission which corresponds to the canonical spectral change from the LH to the HS state when the intensity of the soft component increases. This region corresponds to the minor flaring, major flaring, suppressed and ``post-flaring" states of S08, where the X-ray flux is around or above the transition level and the radio flux varies from below the transition level to above it. Also the HXR changes from an anti-correlation with the radio to a correlation during the major flare events (M99). The SXR variability stays at the level of or decreases slightly from the level of quiescent states. We have further divided the flaring region into three areas to better specify the different X-ray states that occur in this region and to relate them to those presented in S04 and S08. However, we note that the flaring substates appear to be rather continuous instead of being strictly individual states (see Fig. \ref{asmhid}). 

\begin{figure}
\begin{center}
\includegraphics[width=0.5\textwidth]{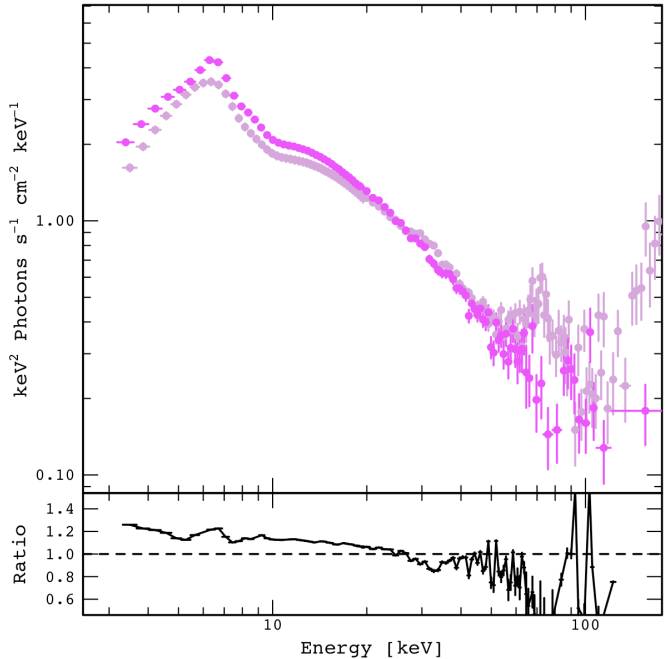}
\caption{\textit{Top:} The average FHXR/major flaring state spectra (dark magenta) with radio flux density above 1 Jy and the average FHXR/minor flaring state spectra (light magenta) with radio flux density below 300 mJy. The sharp spectral feature around 70 keV and increasing flux above 100 keV in FHXR/minor flaring state spectra are most likely due to a calibration issue. \textit{Bottom}: The ratio of the spectra.} \label{fhxr}
\end{center}
\end{figure}

\begin{figure}
\begin{center}
\includegraphics[width=0.5\textwidth]{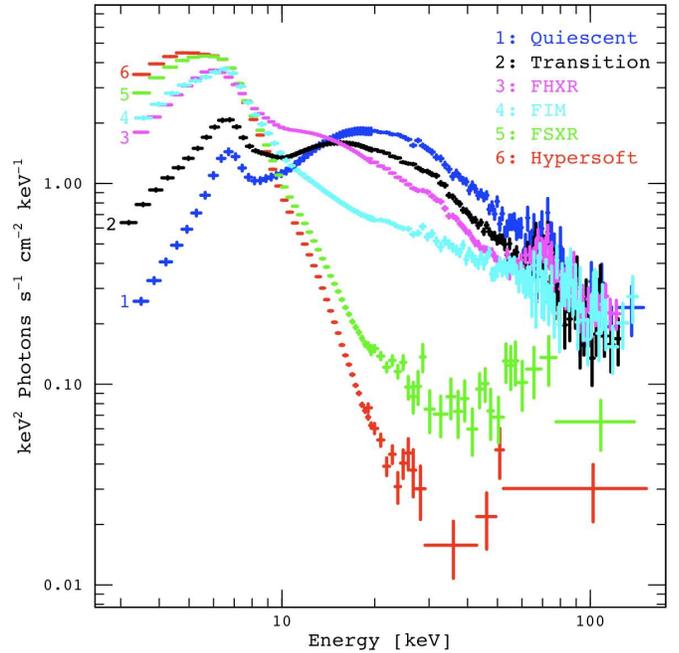}
\caption{The average spectra of X-ray states of Cyg X-3. This figure is similar to Fig. 2 in S04, but for the addition of the hypersoft state. Note also the difference in y-axis. The spectra are coloured and labelled with different colours and numbers according to the legend.} \label{plotall}
\end{center}
\end{figure}

During the FHXR state Cyg X-3 displays various types of flaring. Closer to the transition region Cyg X-3 exhibits minor flaring in the radio but also experiences major flares when the radio emission increases (5--20 Jy). In a normal HID for a black hole system this vertical branch would correspond to the LH state. However, in Cyg X-3, the major flares are spread within the whole flaring region of the HID and they are also found in the FIM and FSXR states. The difference is apparent in the X-ray spectra. The spectrum of the FHXR state is dominated by a thermal component but it also shows a significant amount of HXR. This state represents the spectra of group 3 in S04. To examine any differences in the X-ray spectra associated with different radio behaviour, the composite spectra of the FHXR state are plotted in Fig. \ref{fhxr} for radio flux below 300 mJy (minor radio flaring) and above 1 Jy (major radio flaring). The dramatic difference in radio flux density does not affect the X-ray spectra much. The major radio flaring FHXR spectrum has more flux in the SXR below 25 keV, a slightly steeper power law in the HXR and more flux in the iron line region. The sharp spectral feature around 70 keV in the minor radio flaring FHXR spectrum is most likely due to the \hexte\/ background emission line features. Also, the increasing flux above 100 keV is most likely a calibration issue.  

\subsubsection{Flaring/Intermediate State (FIM)}

This state lies between the areas of FHXR and FSXR in the HID and is characterized by major flaring. The X-ray spectrum is similar to the FHXR spectrum with slightly less contribution to the HXR and corresponds to group 4 spectra in S04. The FIM and FHXR are the states in which Cyg X-3 mainly alternates right after a major flare event.     

\subsubsection{Flaring/Soft X-ray State (FSXR)}

The left vertical branch of the HID consists of two distinct states although they might not be distinguishable at first glance from the X-ray spectra. The whole branch forms the ultrasoft state of S04 (group 5) and quenched state of S08 which is a pre-flare state occurring before a major flare event. However, when taking into account the simultaneous radio data this branch is clearly divided in two parts: the FSXR state and the hypersoft state. The spectrum of the FSXR state is soft without significant HXR emission. However, it shows a weak power law tail extending to higher energies. Very Long Baseline Interferometer (VLBI) observations during one of these types of flares, being a major flare, did not reveal any jet-like structure \citep{mccollough2} indicating that the jet is not yet fully visible in this state or that it is possibly disrupted. The line that demarcates these two states is the so-called \textit{jet line} and most remarkably, unlike transient black hole XRBs, Cyg X-3 crosses this line from right to left (from FSXR to hypersoft) without producing a major radio flare. It is only when it crosses from left to right that a major flare occurs.      

\subsubsection{Hypersoft State}

During the quenched state (S08), the radio emission falls to very low values (10--20 mJy, and often decreases to 1--2 mJy in the Ryle data), the HXR vanishes, and thus the HXR and radio switch from an anti-correlation to a correlation. In this state, the radio flux is well below and the SXR is above the transition level, and it is after Cyg X-3 emerges from this state that a major radio flare occurs. The first two \integral\/ observations of 2007 May are likely representative of this quenched state \citep{beckmann}, during which the X-ray spectrum was very soft (as deduced from the X-ray fluxes in the 2--10 and 20--60 keV bands) with a weak and hard power law tail ($\Gamma$=1.7--1.9).

However, in re-examining the X-ray spectra using our new demarcation technique, we find that there is a group of spectra that are even softer than the FSXR (or the ultrasoft of S04) and in which the power law tail is an order of magnitude fainter than in the quenched state of S08.  This state was previously included in group 5 in S04, but employing the radio fluxes in the HID, we find that this group is in fact distinct from the ultrasoft state. \textit{We denote this new X-ray state of Cyg X-3 the hypersoft state}. 

\section{Discussion}

As mentioned above, Cyg X-3's HID is remarkably similar \textit{in shape} to that of transient black hole XRBs even though it is not strictly speaking a transient source. Despite the similarity, however, Cyg X-3 does not cycle through the HID like other black hole XRBs. Rather, it traces a path from right to left (indicating spectral softening) or then from left to right (spectral hardening), with changes in intensity within each state. By looking at Fig. \ref{asmhid} it appears that the movement across the HID is continuous with no discernible gaps in between. This result differs from that obtained by \citet{smit}, where they found two branches and possibly a third one connecting these two in the HID. However, the discrepancy could arise from the lack of observations or the slightly different hardness used in their work (5.5--19.0 keV/3.5-5.5 keV). From the spectra it appears that there are three major components in play: a thermal disc, a Comptonized thermal component and a Comptonized non-thermal component. When the source moves from right to left the contribution of the thermal disc increases. The changes in this component are simply mapping the accretion on to the compact object. At the same time the HXR component decreases, but there are two effects contributing to this change: the overall HXR flux is dropping and the X-ray spectrum is changing from one that is dominated by thermal Comptonization to one that is dominated by non-thermal Comptonization. The fact that the radio flux density increases when the source enters the flaring region and, at the same time, the HXR switches from an anti-correlation to a correlation during major flares, indicates a relationship between the radio and the non-thermal Comptonization.

\begin{figure}
\begin{center}
\includegraphics[width=0.5\textwidth]{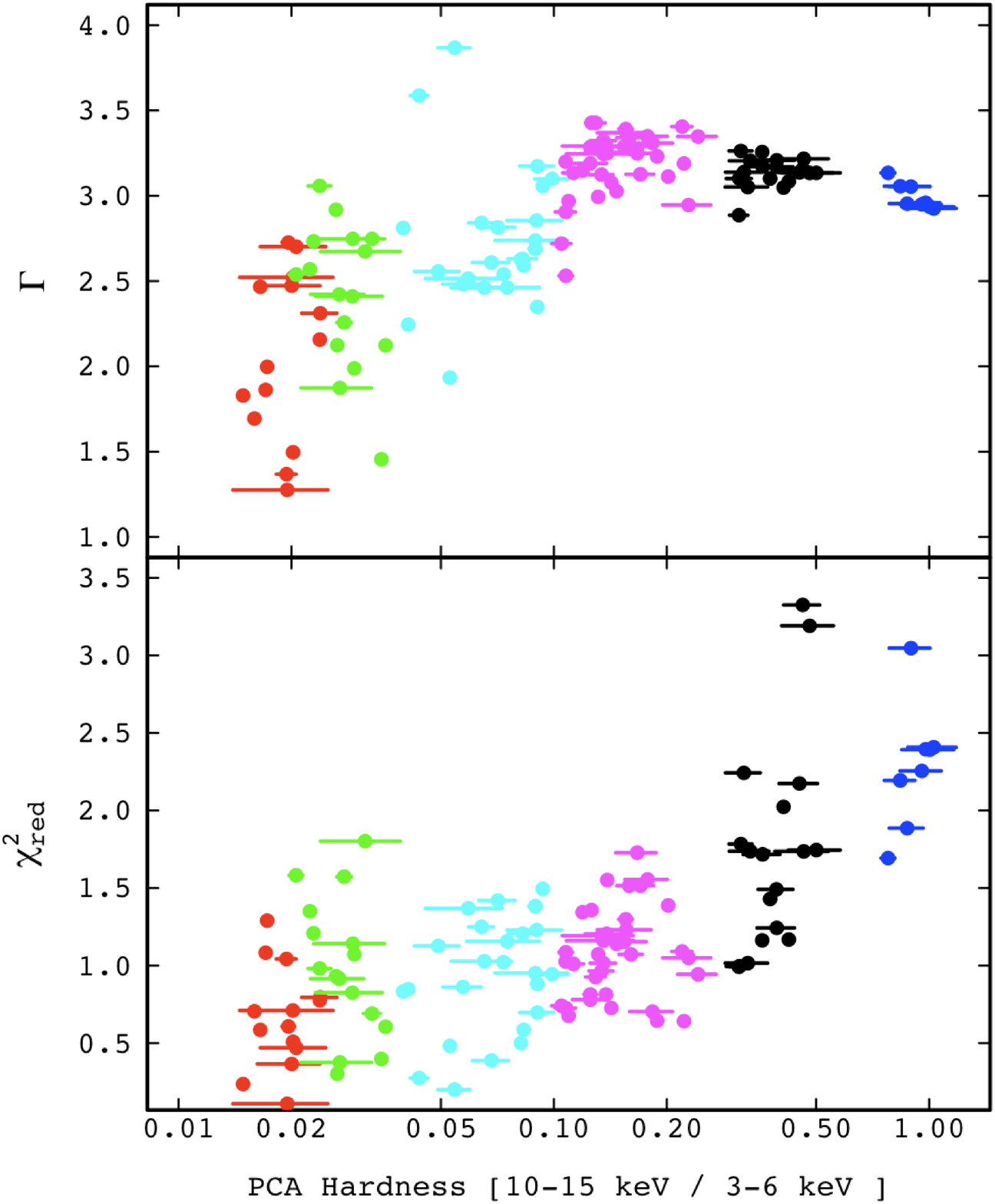}
\caption{Photon indices (top panel) and reduced chi-square values (bottom panel) from fits to pointed \rxte\/ spectra above 20 keV for different SXR hardnesses. The colours are as in Fig. \ref{plotall}. }
\label{curvature}
\end{center}
\end{figure}

To further investigate the thermal/non-thermal Comptonization shift in the source, we have fitted the pointed \rxte\/ spectra above 20 keV with a power law model. The resulting photon index ($\Gamma$) and reduced chi-square values ($\chi_{red}^2$) are plotted in Fig. \ref{curvature}. A larger chi-square value indicates more curvature in the spectra and therefore more thermal Comptonization. It is apparent from the figure that the thermal HXR component is present in the quiescent states and the HXR component is more non-thermal in the flaring states. Also, the photon index changes from $\sim$ 3 in the quiescent states to decreasing steadily from 3 to 1.5 in the flaring states. 

\subsection{Case Study: A Major Radio Flaring Event April-May 2000}

\begin{figure}
\begin{center}
\includegraphics[width=0.5\textwidth]{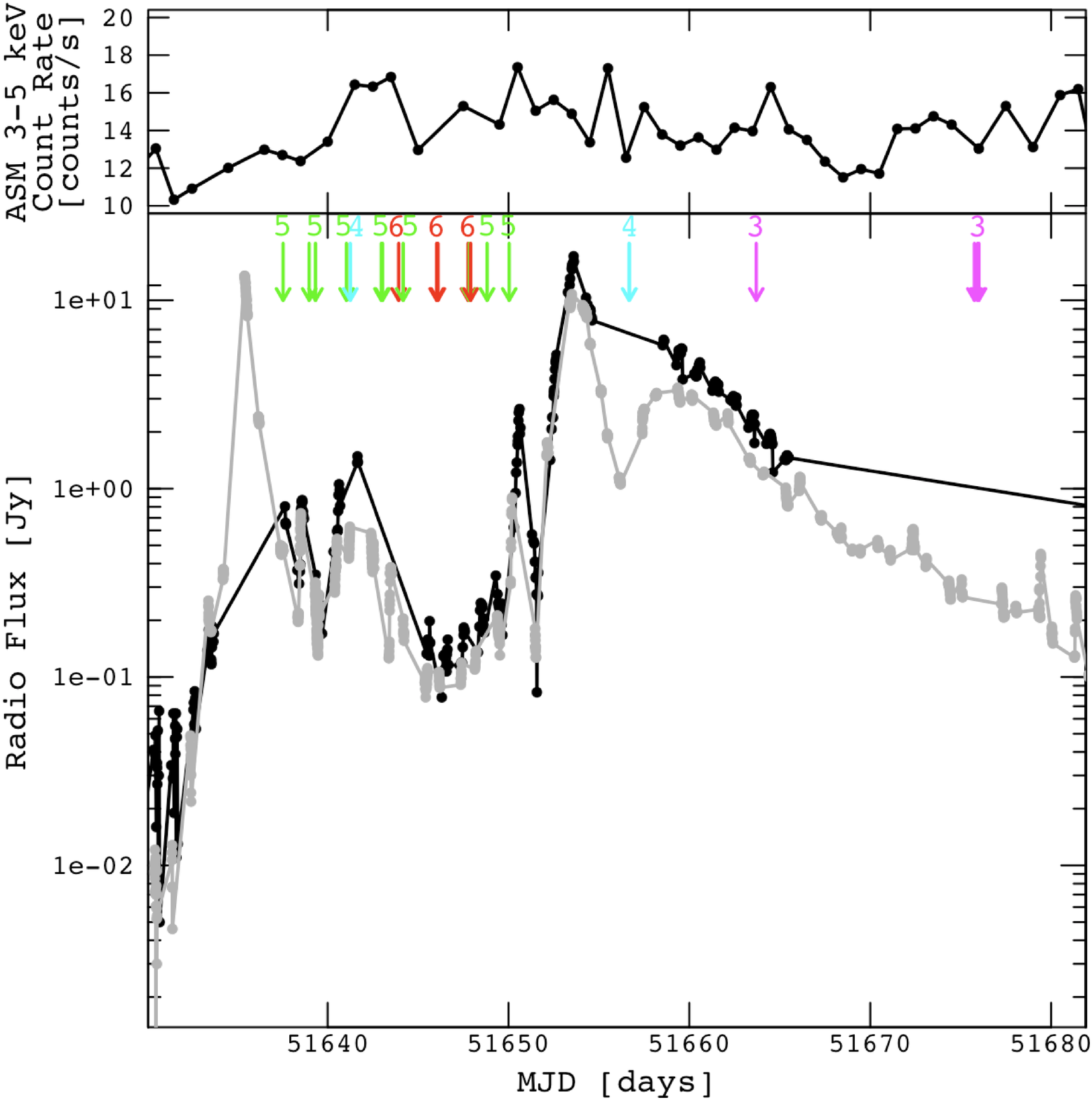}
\caption{The top panel shows the \asm\/ 3--5 keV lightcurve and the bottom panel GBI 8.3 GHz (black) and Ryle 15 GHz radio (grey) lightcurves with simultaneous \rxte\/ pointings (P50062) marked in the upper part of the panel, coloured and numbered as in Fig. \ref{plotall}.} 
\label{flare}
\end{center}
\end{figure}

\begin{figure}
\begin{center}
\includegraphics[width=0.5\textwidth]{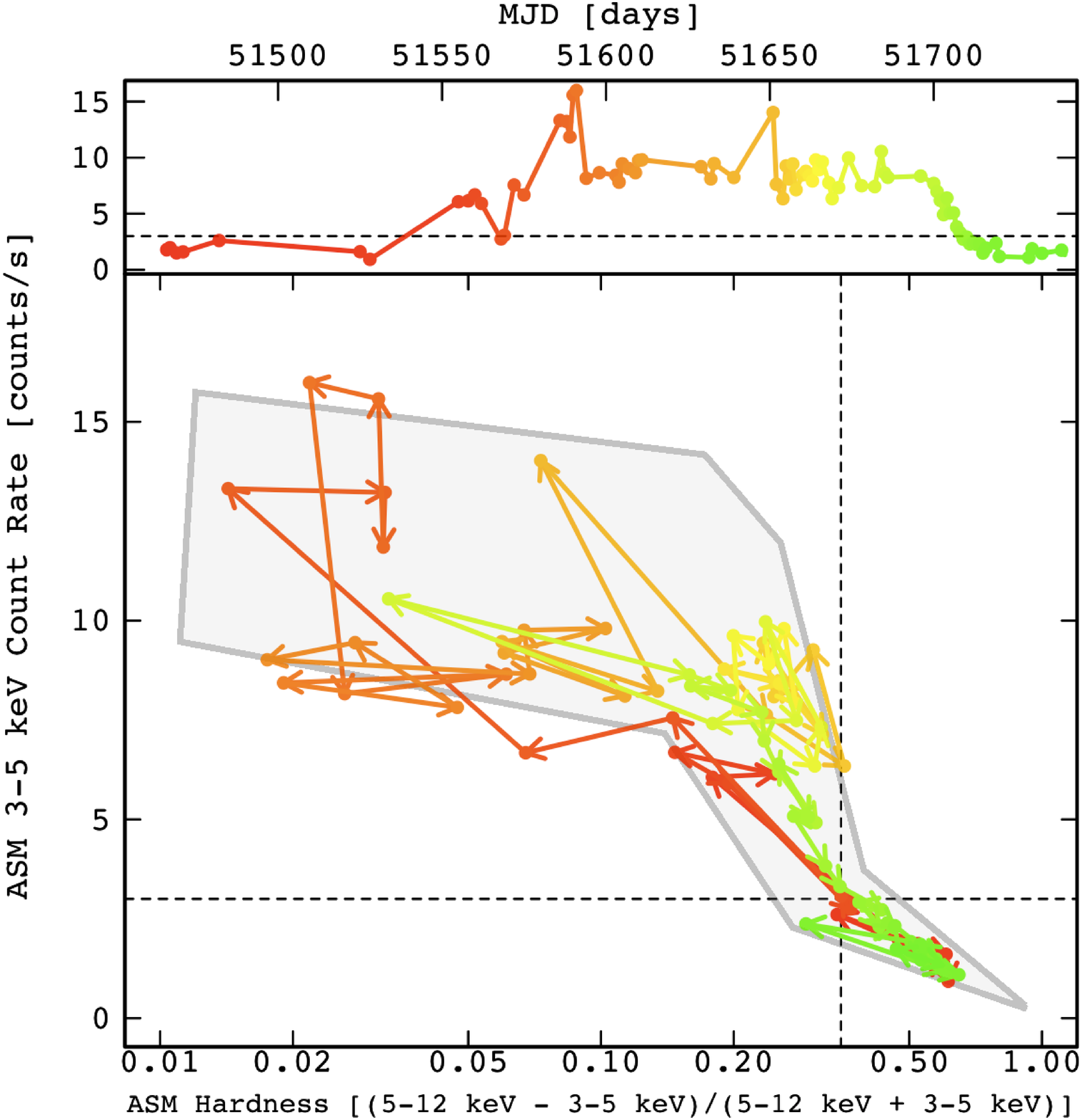}
\caption{The top panel shows the \asm\/ 3--5 keV lightcurve and the bottom panel the evolution of Cyg X-3 in the HID (marked as a grey area) following the system from quiescent state through HYS and flaring states back to quiescence including the major radio flaring event discussed in the Section 5.1. Both panels are color-coded to match each other.} 
\label{hidroutes}
\end{center}
\end{figure}

In Fig. \ref{flare}, a major radio flaring period consisting of two major flares is plotted together with the \rxte\/ pointings (P50062). This same event is part of the state transition curve in Fig. \ref{hidroutes}, which shows the evolution of Cyg X-3 in the HID from quiescent states to flaring states and back. This same period was also analyzed in S08 where they show the succession of the X-ray spectra and corresponding models (in their Fig. 8b). From Figs. \ref{flare} and \ref{hidroutes} we can follow the succession of the radio/X-ray states of Cyg X-3 including periods of major radio flaring. Also in this case Cyg X-3 differs from the transient black hole XRBs that trace their HIDs anti-clockwise. Instead, Cyg X-3 starts in the quiescent state and moves to the hypersoft branch and after a major radio flare enters the FHXR state via the FSXR/FIM states. This flaring episode can repeat as can be seen in Fig. \ref{flare}. Eventually Cyg X-3 will return to the quiescent state. It is hard to pinpoint any particular route that Cyg X-3 follows, except that the different radio/X-ray states are followed in succession. Most likely the spread in the intensity comes from X-ray flaring and hysteresis does not occur in the system. 

From Fig. \ref{hidroutes} we can zoom on the major radio flaring events as shown in Fig. \ref{flare} and see that from the FIM/FSXR state after the first major flare Cyg X-3 enters the hypersoft state just before the second major flare. S08 discussed that during this time of the HXR tail or radio major flare the non-thermal electrons may be accelerated to high enough energies to result in detectable emission in $\gamma$-rays in the GeV/TeV range. Recently, there have been observations of GeV emission before the onset of a major radio flare and when Cyg X-3 exhibited similar properties to the hypersoft state \citep{tavani,corbel}, i.e. high SXR flux and very low or non-existent radio emission, indicating a major ejection of highly relativistic particles which will then most likely form the radio jet a few days later. In this particular case, the decaying of the first major flare leads to a higher radio flux density ($\sim$ 0.1 Jy) that usually is observed together with the hypersoft state. However, it is likely that the system is quenching based on the X-ray state. From the hypersoft state Cyg X-3 enters the FSXR state, and after the peak of the second radio flare, it enters the FIM state. Throughout this progression, the power in the non-thermal HXR increases. During the decline of the second major flare Cyg X-3 enters the FHXR state and thermal Comptonization starts to dominate the HXR spectra. The minor flaring also present in the FHXR state spectra might arise in the aftermath of a major flaring event representing subsequent but smaller ejections. We note that during our case the two major flares present behave very differently with the first flare exhibiting little HXR emission and the second flare exhibiting significantly more (see the discrepancy of hardnesses in Fig. \ref{hidroutes}). A possible mechanism for creating the difference is discussed in Section 5.2.1.       

Cyg X-3 also has the added distinction of exhibiting radio flares \textit{only} when crossing the jet line from left to right (spectral hardening), rather than from right to left (spectral softening) like other black hole XRBs. Fig. \ref{flare} shows that Cyg X-3 produces multiple major radio flares during major flaring periods and additional quenching periods between the flares. However, their detection without accompanying X-ray observations may be difficult because the decaying initial major radio flare dominates the radio emission. 

We also note that the power density spectra (PDS) of Cyg X-3 are not a good indicator of state and that on the short time-scales all the soft state PDS can be fit by a power law with index approximately --2 and that basically there is no power above $\sim$ 1 Hz (e.g. \citealt{axelsson}). This most likely comes from the scattering in the WR wind \citep{zdziarski} and the --2 power law is characteristic of a random walk process likely produced by flares in the X-ray.

\subsubsection{Spectral Analysis}
 
\begin{table*}
\caption{Model parameters for the average spectra of hypersoft state (HYS) as well as the individual consecutive HYS (50062-02-03-02), FSXR (50062-02-07-00), FIM (50062-01-01-00) and FHXR (50062-01-03-00)} spectra. \label{compps}
\begin{center}
\begin{tabular}{lllccccc}
\hline\hline
Component & Parameter & Unit & Avg. HYS & Ind. HYS & Ind. FSXR & Ind. FIM & Ind. FHXR \\
\hline
\texttt{phabs} & $nH_{0}$ & $10^{22}$ & 2f & 2f & 2f & 2f & 2f \\
\texttt{pcfabs} &$nH_{1}$ & $10^{22}$ & 6f & 6f & 6f & 6f & 6f \\
& cov & & 0.9f & 0.9f & 0.9f & 0.9f & 0.9f \\
\texttt{edge} & $E_{Fe}^{H-like}$ & keV & $8.99^{+0.06}_{-0.06}$ & $8.82^{+0.06}_{-0.05}$ & $9.16^{+0.08}_{-0.07}$ & $9.18^{+0.09}_{-0.08}$ & $9.41^{+0.07}_{-0.05}$ \\
& $\tau_{Fe}^{H-like}$ & & $0.35^{+0.02}_{-0.03}$ & $0.30^{+0.02}_{-0.02}$ & $0.25^{+0.02}_{-0.02}$ & $0.18^{+0.02}_{-0.02}$ & $0.26^{+0.02}_{-0.02}$  \\
& $E_{Fe}^{He-like}$ & keV & -- & -- & -- & -- & $7.94^{+0.09}_{-0.09}$ \\
& $\tau_{Fe}^{He-like}$ & & -- & -- & -- & -- & $0.18^{+0.03}_{-0.02}$ \\
\texttt{gauss[Fe]} & $E_{Fe}$ & keV & $6.33^{+0.06}_{-0.07}$ & $6.22^{+0.05}_{-0.06}$ & $6.51^{+0.03}_{-0.03}$ & $6.45^{+0.03}_{-0.03}$ & $6.57^{+0.03}_{-0.03}$ \\
& $\sigma_{Fe}$ & keV & $0.50^{+0.09}_{-0.09}$ & $0.40^{+0.08}_{-0.08}$ & $0.48^{+0.05}_{-0.05}$ & $0.51^{+0.06}_{-0.05}$ & $0.26^{+0.08}_{-0.09}$ \\
& $EW_{Fe}$ & eV & $415^{+73}_{-56}$ & $268^{+44}_{-35}$ & $558^{+52}_{-45}$ & $468^{+45}_{-38}$ & $230^{+25}_{-23}$ \\
\texttt{diskbb} & $kT_{bb}$ & keV & $1.52^{+0.03}_{-0.03}$ & 1.52f & 1.52f & -- & -- \\
& $R_{in}$ & km & $14.1^{+1.2}_{-1.3}$ & $25.9^{+1.1}_{-1.3}$ & $14^{+3}_{-5}$ & -- & -- \\
\texttt{compPS} & $kT_{e}$ & keV & $12.8^{+0.5}_{-0.5}$ & 12.8f & $13.3^{+1.0}_{-1.3}$ & $25.2^{+0.7}_{-0.7}$ & $16.7^{+0.2}_{-0.2}$ \\
& $kT_{bb}$ & keV & $0.75^{+0.03}_{-0.03}$ & $0.88^{+0.03}_{-0.03}$ & $1.03^{+0.02}_{-0.02}$ & $1.09^{+0.01}_{-0.01}$ & $1.05^{+0.01}_{-0.01}$ \\
& $R_{in}$ & km & $74^{+7}_{-5}$ & $83^{+4}_{-2}$ & $64^{+5}_{-6}$ & $73^{+2}_{-1}$ & $74^{+1}_{-2}$ \\
& $\tau$ & & $1.01^{+0.17}_{-0.15}$ & $0.75^{+0.06}_{-0.05}$ & $1.1^{+0.2}_{-0.1}$ & $0.93^{+0.06}_{-0.05}$ & $>2.89a$ \\
& $\Gamma$ & & 2f & 2f & 2f & $4f$ & -- \\ 
\hline
\texttt{Fluxes} & $F_{bol,abs}^{b}$ & $10^{-9}$ erg cm$^{-2}$ s$^{-1}$ & $9^{+2}_{-1}$ & $28^{+2}_{-2}$ & $22^{+5}_{-5}$ & $44^{+2}_{-2}$ & $42^{+1}_{-2}$ \\ 
& $F_{bol}^{c}$ & $10^{-9}$ erg cm$^{-2}$ s$^{-1}$ & $13^{+2}_{-2}$ & $40^{+3}_{-3}$ & $31^{+8}_{-7}$ & $55^{+3}_{-2}$ & $51^{+2}_{-2}$ \\
& $L_{bol}^{d}$ & $10^{38}$ erg s$^{-1}$ & $1.2^{+0.2}_{-0.2}$ & $3.9^{+0.3}_{-0.3}$ & $3.0^{+0.7}_{-0.7}$ & $5.3^{+0.2}_{-0.2}$ & $5.0^{+0.2}_{-0.2}$ \\
& $F_{comp}/F_{bol}$ & & 0.44 & 0.41 & 0.75 & 0.96 & 0.98 \\
& $F_{disk}/F_{bol}$ & & 0.51 & 0.55 & 0.19 & -- & -- \\
& $F_{line}/F_{bol}$ & & 0.05 & 0.04 & 0.06 & 0.04 & 0.02 \\
\hline
\texttt{Confidence} & $\chi^{2}_{red}$/d.o.f. & & 1.02/41 & 0.98/32 & 1.24/36 & 1.04/74 & 1.24/70 \\
\hline
\end{tabular}
\end{center}
\begin{list}{}{}
\item[$^{\mathrm{a}}$] Only the lower limit is shown because the optical depth is pegged to the highest value.
\item[$^{\mathrm{b}}$] Absorbed, bolometric flux (3--200 keV) of the model normalized to the \pca\/ data.
\item[$^{\mathrm{c}}$] Unabsorbed, bolometric flux (3--200 keV) of the model normalized to the \pca\/ data.
\item[$^{\mathrm{d}}$] Unabsorbed, bolometric luminosity (3--200 keV) of the model assuming a distance 9 kpc to the source normalized to the \pca\/ data.
\item[$^{\mathrm{f}}$] Frozen in the fits.
\end{list}
\end{table*}

\begin{figure}
\begin{center}
\includegraphics[width=0.5\textwidth]{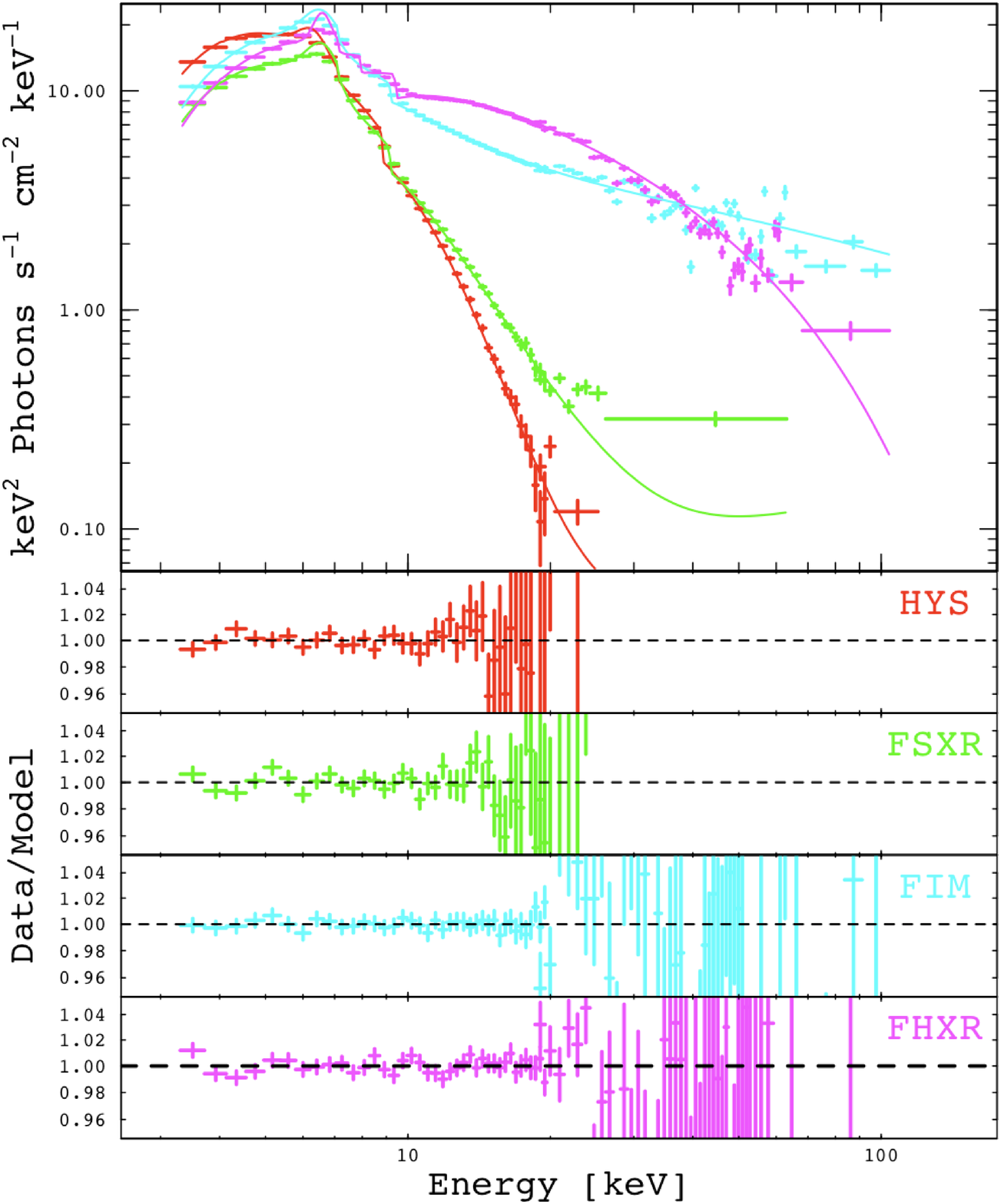}
\caption{The individual and consecutive data of the hypersoft (red), FSXR (green), FIM (cyan) and FHXR (magenta) states with best fit models. See details in the text and parameters in Table \ref{compps}.} 
\label{hysmodel}
\end{center}
\end{figure}

In the following we fit a model to the average hypersoft spectra and consequent individual HYS, FSXR, FIM and FHXR spectra during the second major flaring event. 

With the exception of the hypersoft state the other spectral states of S04 have been discussed in S04, \citet{hjalmarsdotter1} and \citet{hjalmarsdotter2}. In the following we will examine the hypersoft state and compare it to the flaring states. We fit the average hypersoft spectra as well as individual, consecutive hypersoft (50062-02-03-02), FSXR (50062-02-07-00), FIM (50062-01-01-00) and FHXR (50062-01-03-00) spectra during a major radio flare event, this being the only major flare event in the data set where there is \rxte\/ coverage and showing the succession of these states. The model includes neutral absorption and partially covered absorption, a gaussian line for the iron complex (the dominant line being the He-like iron line, but including contributions also from the H-like line, see e.g. \citealt{paerels}) and ionized iron edges (contributions most likely from the H-like and He-like edges). We model the continuum with \compps\/, which is a hybrid Comptonization model \citep{svensson} including multicolour disc black body emission as seed photons for the Comptonization process. Additionally, we include an unscattered disc component in the model, but the best fits appear to suppress this component from the FIM and FHXR spectra and therefore it is excluded from those.

The absorption components are frozen to average values found by Koljonen et al. (in prep.) based on multiple \swiftxrt\/ spectra, as these provide better coverage below 3 keV as opposed to the \rxtepca\/. The absorption edges of highly ionized iron (H- and He-like) improve the fits substantially as is noted in \citet{hjalmarsdotter1} and therefore have been included in the model. Cyg X-3 harbors a large number of emission lines that are especially prominent in the high-resolution energy spectra, e.g. \citet{paerels}, but are also visible in the \swiftxrt\/ data. These lines arise in the stellar wind of the WR companion ionized by the intense X-ray emission from the compact object. However, due to the poor spectral resolution of the \pca\/, only the iron complex region is prominent in the spectra ($E_{Fe}\sim6.5$ keV in the fits). For \compps\/, we use the following settings: the electron distribution function is hybrid with a low temperature Maxwellian plus a power-law tail (Lorenz factor extending from $\gamma_{min}=1.3$ to $\gamma_{max}=1000$) for the HYS, FSXR and FIM spectra. As evidenced in Fig. \ref{curvature} the FHXR spectrum shows curvature and therefore we fit only the Maxwellian distribution to the FHXR spectrum. The electron power law index is not well constrained in the fits, so we have frozen that to $\Gamma=2$ for the HYS and FSXR spectra, a typical value for soft states \citep{hjalmarsdotter1}, and $\Gamma=4$ for the FIM spectrum. The seed photons arise in a multicolour disc black body (\diskbb\/, \citealt{mitsuda}) with inner disc temperature as a characterizing parameter. We also consider unscattered disc black body emission and therefore include also an extra disc component to the model. We do not include any reflection component in the model. The inclination of the accretion disc to the line-of-sight is uncertain and dependent on the nature of the compact object, and in the wind-fed binary system it does not even need to lie in the orbital plane \citep{vilhu2}. However, we find that exchanging from sphere-approximated Comptonization to slab Comptonization does not improve the fits. To summarize, we have only the scattered and unscattered inner disc temperatures, the electron temperature and the optical depth of the electron cloud as free parameters in addition to the gaussian line and the iron edges. We also allow for an intercalibration factor between \pca\/ and \hexte\/ data correcting for calibration differences. The best-fitting parameters can be found in Table \ref{compps} and we see that this model fit the spectra well in the HYS and FIM states ($\chi^{2}_{red}$ $\sim$ 1.0) and acceptably in the FSXR and FHXR states ($\chi^{2}_{red}$ $\sim$ 1.2). 

We find from the modelling that the HYS and flaring states can be fit with a model consisting of an absorbed disc black body and hybrid Comptonization including spectral features from iron emission lines and edges. From Fig. \ref{hysmodel} and the fit results we see that for the HYS, FSXR and FIM states, non-thermal Comptonization is needed to account for the HXR tail, and that the FHXR can be fit with only thermal Comptonization. We note that this model is not adequate for fitting the quiescent state spectra and most likely reflection \citet{szostek2} and/or Compton downscattering  \citet{zdziarski} takes place in these states. From the succession of the spectral states we notice that the soft seed photon temperature for the Comptonization remains at $kT_{bb}\sim1$ keV and the optical depth of the plasma remains at $\tau\sim1$ except when moving into the FHXR state where it pegs to the upper limit. However, high plasma optical depths of $\tau\sim3-6$ are suggested to explain the hard spectra of Cyg X-3 in \citet{hjalmarsdotter2} and \citet{zdziarski}. We find the electron temperature to be rather soft, $kT_{e}\sim13-16$ keV, except in the FIM state where the best-fitting value is found to be $\sim25$ keV. The FIM state is also the state where the highest radio flux densities are found, so this might indicate that the radio jet has an effect on the Comptonizing electron temperatures. On the other hand, in our case study the FIM state occurs after a $\sim$ 10 Jy radio flare and the FHXR state after a $\sim$ 2 Jy flare (see Ryle data in Fig. \ref{flare}), which might have an impact on the order of these states occurring after a major radio flare. One would expect the FHXR state to occur right after the major flare peak exhibiting higher optical depth and therefore coming from a more compact area, likely closer to the base of the jet. As the flare evolves and becomes more optically thin one would expect the system to enter the FIM state exhibiting a more power law-like tail.   

We also note that for the HYS and FSXR states an unscattered disc component is required to achieve acceptable fits. The average HYS spectrum yields an inner disc temperature $kT_{bb}=1.52$ keV and this value is frozen for the individual spectral fits in order to gauge the level of unscattered disc flux in each state. We find that this disc component accounts for 50\% of the X-ray flux in the HYS state and drops to 20\% in the FSXR state; for the FIM and FHXR states all the disc photons appear to be scattered. We note that locking the two black body temperatures to each other in the model does not lead to an acceptable fit. The discrepancy of the inner disc temperature and the soft seed photon temperature for the Comptonization component in the HYS and FSXR state may or may not be a real effect, though it is required for acceptable fits. It is suggested in \citet{szostek2} that there is strong wind emission present below 1 keV in the system. This emission might be the source of the soft seed photons that are Comptonized and causing the scattered disk temperature of $\lesssim$ 1 keV in the model throughout the states (the wind being present at all times in the system). The unscattered disk component in the HYS and FSXR states would represent the accretion disk either forming during these states and launched in the jet (as this component is not present in the FIM and FHXR states), or then remaining invisible to us due to the WR wind. During a major flare event the disc would become visible as the jet plows through the wind. Of course, only extensive SEDs will enable us to determine the actual contribution to the X-ray emission of the relativistic jets, especially during the major flaring periods.

\subsection{Different Emission Scenarios in Cyg X-3}

In the following we discuss briefly the different methods of producing the HXR emission in Cyg X-3. However, more studies are needed for determining the exact process(es) giving rise to the HXR emission.

\subsubsection{Disrupted/Attenuated Jet}

The HID in Fig. \ref{hidradio} indicates that flares with more prominent HXR emission are brighter in the radio than those with less contribution from the HXR (although this might simply be a selection effect). This could be a manifestation of a disrupted or an attenuated jet (\citealt{perucho}, for an actual observation of a disrupted jet in AGN see e.g. \citealt{evans}), i.e. a jet with a less significant HXR component, being therefore less energetic, may not be capable of plowing through the stellar wind of the companion. This effect is clearly present in our case study (Section 5.1) showing a major flare with very low HXR emission followed by a major flare with significant HXR emission. For this to be plausible, then the HXR emission must originate in the jet itself via synchrotron or synchrotron self-Compton (SSC), see e.g. \citet{markoff,bosch}. However, the \asm\/ HID appears to be continuous in the flaring region as well, so the jet in Cyg X-3 would then go from being disrupted to unimpeded. Also, if the jet composition were hadronic, there might be hadronic jet-wind interactions producing the high-energy $\gamma$-ray emission and possibly also a significant amount of neutrinos \citep{romero}. However, the variable and steep $\gamma$-ray spectrum of Cyg X-3 suggests an electron origin, especially if the electrons were located away from the accretion disc.    

\subsubsection{Comptonization processes}

Typically, the HXR emission in XRBs and microquasars is modelled by the Compton upscattering of a soft photon source, i.e. an accretion disc or some other SXR plasma component, by high-energy electrons in a corona surrounding the compact object. The nature of this corona is much debated and it could be e.g. patchy high-energy regions above the accretion disc (e.g. \citealt{zdziarski1}), the base of the relativistic jet (e.g. \citealt{markoff}), or an ADAF-like hot accretion flow (e.g. \citealt{esin}). In the case of microquasars the relativistic particles could be created in or near the accretion disc and then be ejected through the jet. The HXR emission could then arise from the Compton upscattering of the soft accretion disc photons by these relativistic particles and the radio emission from the synchrotron radiation by those ejected in the jet.

However, because the system is embedded in the dense stellar wind of the companion WR star, it brings another level of complexity to the system. In addition to the accretion disc around the compact object the surrounding plasma also affects the HXR photons in several ways including absorption, reflection \citep{hjalmarsdotter1} and Compton downscattering \citep{zdziarski}. The latter would occur in a cold ($\sim 3 \, {\rm keV}$) plasma cloud that would reduce the HXR cut-off to $\sim20 \, {\rm keV}$ \citep{hjalmarsdotter1} as observed from Cyg X-3 as opposed to the usual value of $\sim100 \, {\rm keV}$ observed in other XRBs, especially describing the quiescent X-ray spectra of Cyg X-3. This effect would be most likely present also in the flaring states, however since the thermal HXR emission changes to non-thermal emission in the flaring states as discussed above the effect is not so prominent in the X-ray spectra compared to the quiescent states.    

\subsubsection{Shock-in-jet}

Particle acceleration scenarios in the shock along the jet are very viable and offer an alternative way of producing the relativistic electrons as opposed to particle creation in or near the accretion disc. The so-called shock-in-jet model has been invoked successfully to describe Cyg X-3's radio behaviour during major flaring periods \citep{lindfors} and low-level activity \citep{millerjones}. These shocks would provide the non-thermal power law emission and the link to the radio and HXR during flaring episodes. The relativistic particles would be created in the internal shocks in the jet, and the HXR emission could arise either from synchrotron emission from the electrons in the jet or Comptonization of the disc photons by the high-energy electrons in the shock, depending on the location of the shock and emission regions. In fact, Comptonization of the disc photons by this shock could be in the form of bulk-motion Comptonization (BMC, e.g. \citealt{titarchuk}; however in that paper the BMC is related to the accretion near the compact object and not to the outflow), since the shock might move along the jet with a significant fraction of the speed of light. This non-thermal shock emission may very well also be the source of the recently observed very high energy $\gamma$-ray emission.

If the low-level radio activity can be modelled also with the shock-in-jet model, can we apply this to the quiescent X-ray spectra? In fact, the Cyg X-3 quiescent state X-ray spectra can be modelled with an almost purely non-thermal electron distribution which contributes to the non-thermal Comptonization of SXR photons \citep{hjalmarsdotter1}. In this case the peak of the Comptonized spectrum is no longer determined by the maximum electron temperature, but rather the energy where the majority of electrons are injected explaining the unusually low cut-off in the Cyg X-3 quiescent X-ray spectra. In this way the shock-in-jet model would provide the necessary elements of modelling both the quiescent and flaring X-ray and radio spectra.    

\section{Summary}

We have reviewed the main X-ray and radio characteristics of Cyg X-3 and noted the complexity of the X-ray spectra and the tight interplay between the radio and X-rays in the source. We have investigated the hardness-intensity diagram (HID) of Cyg X-3 and shown how it displays both similarities to and differences with black hole transient HIDs during outburst. However, the possible black hole nature of the compact object is not resolved by this study, since the specific black hole signature, hysteresis, is not present or it is smeared out in the HID. When combined with the radio data, the radio/X-ray states are clearly revealed in the HID and we can see that the major radio flares in Cyg X-3 occur in a wide range of X-ray hardness. In addition, we categorize a new X-ray state, the hypersoft state, that has a hint of a hard energy tail and is linked to the quenched radio state. When the source exhibited similar conditions to the hypersoft state there have been detections of $>$100 MeV flux, suggesting that something extremely energetic is occurring in the system during this state. When Cyg X-3 crosses the jet line of the HID from this hypersoft state to a flaring state, a major radio flare occurs. However, no major flares have been observed when the system crosses the line in other direction. We have discussed the possible origin of the varying HXR emission and concluded that it consists of thermal Comptonization and non-thermal Comptonization that are linked to the radio emission. This link could be explained by the creation of high-energy particles in or near the accretion disc some of which are ejected in the jet and some Comptonize the SXR photons, or shocks traveling in the jet causing the observed non-thermal emission and possibly the recently observed $\gamma$-ray emission. 

Unsolved issues raised by this work are: 

(a) What is the reason for the major radio flares occurring in a wide range of X-ray hardness? Is it due to jet disruption by the stellar wind or some variable HXR producing emission mechanism?

(b) What is(are) the exact process(es) producing the HXR emission? Is it due to Compton upscattering, synchrotron emission from the electrons in the jet, or a shock-in-jet scenario?

(c) What is the physical scenario during the hypersoft state in the system, where the HXR and radio emission are at a minimum, but very energetic ($> $100 MeV) photons have been observed?

Future monitoring programs designed to catch Cyg X-3 in the hypersoft state will be helpful in alerting the multiwavelength community for setting up a comprehensive campaign. Of particular importance are the radio and hard X-rays as these will truly inform us that the hypersoft state is occurring. Follow-up observations in radio imaging alongside spectral energy distributions up to $\gamma$-rays give us a better understanding on the emission processes throughout major flaring events.  

\section*{Acknowledgments}

The authors thank the referee for careful reading of the manuscript and useful suggestions that improved the paper. The authors also thank Osmi Vilhu for enlightening discussions. KIIK gratefully acknowledges a grant from Jenny ja Antti Wihurin s\"a\"ati\"o and Acadamy of Finland grant (project num. 125189). DCH acknowledges Academy of Finland grant (project num. 212656). MLM wished to acknowledge support from NASA under
grant/contract NNG06GE72G, NNX06AB94G and NAS8-03060. ST acknowledges support from Russian Foundation for Basic Research,
grant 08-02-00504-a. This research has made use of data obtained from the High Energy Astrophysics Science Archive Research center (HEASARC), provided by NASA's Goddard Space Flight center.

\appendix

\section[]{Observation log}

\begin{table*}
\caption{Observations in chronological and obsid order. Along with the observation times, the \rxtepca\/ and \rxtehexte\/ exposures, the radio/X-ray state of the system and mean radio flux (during or near the \rxte\/  observations) are marked. In the radio/X-ray state column \textit{Q} stands for quiescent state, \textit{T} for transition, \textit{FHXR} for flaring/hard X-ray, \textit{FIM} for flaring/intermediate, \textit{FSXR} for flaring/soft X-ray and \textit{HYS} for hypersoft state. See more information from the classification of the states in the Section 4. In the radio flux column \textit{Ry} stands for Ryle 15 GHz, \textit{G} for GBI 8.3 GHz and \textit{RA} for RATAN-600 11.2 GHz. ``--`` stands for no radio observations.}
\begin{center}
\begin{tabular}{|lccccccc}
\hline
\hline
No. & ObsID & Date & MJD & PCA & HEXTE & X-ray/ & Radio \\
& & & Interval & Exp. & Exp. & Radio & Flux \\
& & [yy/mm/dd] & [d] & [ks] & [ks] & State & [Jy] \\
\hline
1 & 10126-01-01-00 & 96/08/24 & 50319.455--50319.690 & 12.4 & 3.7 & T & 0.042 (Ry) \\
2 & 10126-01-01-010 & 96/08/26 & 50321.150--50321.431 & 14.5 & 4.2 & T & 0.028 (Ry) \\
3 & 10126-01-01-01 & 96/08/26 & 50321.456--50321.631 & 9.9 & 3.0 & T & 0.028 (Ry) \\
4 & 10126-01-01-020 & 96/08/27 & 50322.257--50322.498 & 13.5 & 3.9 & T & 0.017 (Ry) \\
5 & 10126-01-01-02 & 96/08/27 & 50322.523--50322.762 & 11.7 & 3.6 & T & 0.017 (Ry) \\
6 & 10126-01-01-04 & 96/08/28 & 50323.663--50323.831 & 7.7 & 2.3 & T & 0.046 (Ry) \\
7 & 10126-01-01-03 & 96/08/29 & 50324.665--50324.765 & 5.4 & 1.7 & T & 0.035 (Ry) \\
8 & 10126-01-01-05 & 96/08/30 & 50325.665--50325.822 & 6.9 & 2.1 & T--FHXR & 0.025 (Ry) \\
\hline
9 & 20099-01-01-00 & 97/02/16 & 50495.017--50495.308 & 14.7 & 4.6 & FHXR & 0.063 (G) \\ 
10 & 20099-01-01-010 & 97/02/21 & 50500.018--50500.258 & 13.4 & 4.3 & FHXR & 0.088 (G) \\
11 & 20099-01-01-01 & 97/02/21 & 50500.284--50500.510 & 12.2 & 3.6 & FHXR & 0.088 (G) \\
12 & 20099-01-01-020 & 97/02/21 & 50500.765--50501.059 & 13.8 & 4.3 & FHXR & 0.086 (G) \\
13 & 20099-01-01-02 & 97/02/22 & 50501.084--50501.126 & 3.4 & 1.1 & FHXR & 0.097 (G) \\
14 & 20099-02-01-00 & 97/09/26 & 50717.321--50717.497 & 10.3 & 3.2 & T & 0.205 (G) \\
\hline
15 & 20101-01-01-00 & 97/06/05 & 50604.751--50604.845 & 5.6 & 1.8 & FSXR--FIM & 0.140 (G) \\
16 & 20101-01-02-00 & 97/06/07 & 50606.005--50606.100 & 5.7 & 1.7 & FSXR--FIM & 0.131 (G) \\
17 & 20101-01-03-00 & 97/06/10 & 50609.540--50609.644 & 6.1 & 1.9 & FIM & 0.259 (G) \\
18 & 20101-01-05-00 & 97/06/13 & 50612.472--50612.575 & 5.7 & 1.7 & FIM--FHXR & 1.521 (G) \\ 
19 & 20101-01-06-00 & 97/06/17 & 50616.762--50616.851 & 5.4 & 1.6 & FIM & 0.211 (G) \\ 
20 & 20101-01-04-00 & 97/06/19 & 50618.478--50618.578 & 5.7 & 1.8 & FHXR & 0.449 (G) \\ 
21 & 20101-01-07-00 & 97/06/25 & 50624.723--50624.835 & 5.3 & 1.5 & FIM & 0.430 (G) \\ 
22 & 20101-01-08-00 & 97/07/03 & 50632.689--50632.799 & 6.6 & 2.0 & FHXR & 0.069 (G) \\ 
23 & 20101-01-09-00 & 97/07/23 & 50652.629--50652.724 & 5.6 & 1.6 & Q & 0.026 (G) \\ 
24 & 20101-01-10-00 & 97/08/01 & 50661.697--50661.791 & 5.7 & 1.8 & Q & 0.070 (G) \\
\hline
25 & 30082-04-01-00 & 98/05/16 & 50949.628--50949.717 & 5.2 & 1.5 & Q & 0.100 (G) \\
26 & 30082-04-02-00 & 98/05/17 & 50950.949--50951.050 & 5.3 & 1.7 & Q & 0.063 (G) \\
27 & 30082-04-03-00 & 98/05/18 & 50951.692--50951.783 & 5.4 & 1.5 & Q & 0.069 (G) \\
28 & 30082-04-04-00 & 98/05/19 & 50952.690--50952.783 & 5.5 & 1.6 & Q & 0.065 (G) \\
29 & 30082-04-05-00 & 98/05/20 & 50953.691--50953.783 & 5.4 & 1.6 & Q & 0.056 (G) \\
30 & 30082-04-06-00 & 98/05/21 & 50954.880--50954.983 & 5.8 & 1.9 & Q & 0.044 (G) \\
\hline
31 & 40422-01-01-00 & 99/08/14 & 51404.439--51404.633 & 10.3 & 3.3 & FHXR & 0.068 (G) \\
\hline
32 & 40061-01-01-00 & 00/02/11 & 51585.311--51585.340 & 2.4 & 0.8 & HYS & 0.110 (G) \\
33 & 40061-01-02-00 & 00/02/12 & 51586.378--51586.401 & 2.0 & 0.7 & HYS & 0.292 (G) \\
34 & 40061-01-03-00 & 00/02/13 & 51587.235--51587.258 & 2.0 & 0.7 & HYS & 0.068 (G) \\
35 & 40061-01-04-00 & 00/02/14 & 51588.233--51588.257 & 2.0 & 0.7 & HYS & 0.106 (G) \\
36 & 40061-01-04-01 & 00/02/14 & 51588.305--51588.332 & 2.3 & 0.8 & HYS & 0.062 (G) \\
37 & 40061-01-04-02 & 00/02/14 & 51588.374--51588.399 & 2.1 & 0.7 & HYS & 0.074 (G) \\
38 & 40061-01-05-00 & 00/02/15 & 51589.231--51589.254 & 1.9 & 0.6 & HYS & 0.172 (G) \\
39 & 40061-01-05-01 & 00/02/15 & 51589.301--51589.331 & 2.4 & 0.8 & HYS & 0.234 (G) \\ 
40 & 40061-01-06-00 & 00/02/16 & 51590.229--51590.255 & 2.2 & 0.7 & FSXR & 0.274 (G) \\
41 & 40061-01-06-01 & 00/02/16 & 51590.298--51590.326 & 2.4 & 0.8 & FSXR & 0.323 (G) \\ 
42 & 40061-01-07-00 & 00/02/16 & 51590.953--51591.059 & 6.7 & 2.2 & FSXR & 0.280 (G) \\ 
43 & 40061-02-01-00 & 00/02/18 & 51592.085--51592.186 & 6.0 & 2.0 & HYS & 0.028 (G) \\
44 & 40061-02-02-01 & 00/02/19 & 51593.291--51593.315 & 2.1 & 0.7 & HYS & 0.019 (G) \\
45 & 40061-02-03-00 & 00/02/20 & 51594.150--51594.252 & 5.8 & 1.9 & HYS & -- \\
46 & 40061-02-04-00 & 00/02/20 & 51594.943--51595.077 & 6.4 & 2.2 & HYS & -- \\
47 & 40061-02-05-00 & 00/02/22 & 51596.145--51596.247 & 5.7 & 1.8 & HYS & -- \\
48 & 40061-02-06-00 & 00/02/22 & 51596.939--51597.045 & 6.4 & 2.2 & HYS & -- \\
\hline
\end{tabular}
\end{center}
\label{list1}
\end{table*}%

\begin{table*}
\contcaption{}\begin{center}
\begin{tabular}{|lccccccc}
\hline
\hline
No. & ObsID & Date & MJD & PCA & HEXTE & X-ray/ & Radio \\
& & & Interval & Exp. & Exp. & Radio & Flux \\
& & [yy/mm/dd] & [d] & [ks] & [ks] & State & [Jy] \\
\hline
49 & 50062-02-01-00 & 00/04/03 & 51637.536--51637.621 & 4.3 & 1.4 & FSXR & 0.703 (G) \\ 
50 & 50062-02-02-00 & 00/04/04 & 51638.978--51639.067 & 4.8 & 1.6 & FSXR & 0.521 (G) \\ 
51 & 50062-02-02-01 & 00/04/05 & 51639.327--51639.349 & 1.8 & 0.6 & FSXR & 0.349 (G) \\ 
52 & 50062-02-03-00 & 00/04/07 & 51641.042--51641.135 & 4.8 & 1.6 & FSXR & 0.898 (G) \\ 
53 & 50062-02-03-01 & 00/04/07 & 51641.252--51641.276 & 2.0 & 0.7 & FIM & 1.428 (G) \\
54 & 50062-02-04-00 & 00/04/08 & 51642.967--51642.995 & 2.4 & 0.8 & FSXR & -- \\ 
55 & 50062-02-04-01 & 00/04/09 & 51643.035--51643.058 & 2.0 & 0.6 & FSXR & -- \\ 
56 & 50062-02-05-00 & 00/04/09 & 51643.928--51643.991 & 2.9 & 0.9 & HYS & -- \\
57 & 50062-02-05-01 & 00/04/10 & 51644.173--51644.267 & 4.4 & 1.5 & FSXR & -- \\ 
58 & 50062-02-06-00 & 00/04/12 & 51646.027--51646.049 & 1.9 & 0.6 & HYS & 0.078 (G) \\
59 & 50062-02-06-01 & 00/04/12 & 51646.096--51646.124 & 2.5 & 0.8 & HYS & 0.078 (G) \\
60 & 50062-02-03-02 & 00/04/13 & 51647.764--51647.786 & 1.9 & 0.6 & HYS & 0.174 (G) \\ 
61 & 50062-02-06-02 & 00/04/13 & 51647.837--51647.854 & 1.5 & 0.4 & FSXR & 0.176 (G) \\ 
62 & 50062-02-06-03 & 00/04/13 & 51647.905--51647.923 & 1.5 & 0.5 & HYS & 0.169 (G) \\
63 & 50062-02-07-00 & 00/04/14 & 51648.814--51648.851 & 3.2 & 1.1 & FSXR & 0.188 (G) \\ 
64 & 50062-02-08-00 & 00/04/16 & 51650.037--51650.108 & 3.0 & 1.0 & FSXR & 0.623 (G) \\ 
65 & 50062-01-01-01 & 00/04/22 & 51656.668--51656.686 & 1.6 & 0.5 & FIM & 0.955 (RA) \\ 
66 & 50062-01-01-00 & 00/04/22 & 51656.726--51656.829 & 6.1 & 2.0 & FIM & -- \\ 
67 & 50062-01-03-00 & 00/04/29 & 51663.700--51663.792 & 5.3 & 1.7 & FHXR & 2.203 (G) \\
68 & 50062-01-04-02 & 00/05/11 & 51675.760--51675.770 & 0.9 & 0.3 & FHXR & 0.369 (RA) \\
69 & 50062-01-04-01 & 00/05/11 & 51675.942--51675.963 & 1.8 & 0.5 & FHXR & 0.369 (RA) \\
70 & 50062-01-04-00 & 00/05/12 & 51676.009--51676.039 & 2.6 & 0.8 & FHXR & 0.394 (RA) \\
\hline
71 & 70062-01-01-00 & 02/10/19 & 52566.837--52566.856 & 1.4 & 0.5 & FHXR & 0.110 (Ry) \\
72 & 70062-01-01-01 & 02/10/19 & 52566.905--52567.053 & 5.0 & 1.7 & FHXR & 0.093 (Ry) \\
73 & 70062-01-02-00 & 02/10/20 & 52567.824--52567.844 & 1.5 & 0.5 & FHXR & 0.112 (Ry) \\
74 & 70062-01-02-01 & 02/10/20 & 52567.892--52567.975 & 3.0 & 1.0 & FHXR & 0.104 (Ry) \\
75 & 70062-01-02-02 & 02/10/20 & 52567.997--52568.041 & 2.0 & 6.7 & FHXR & 0.133 (Ry) \\
76 & 70062-01-03-01 & 02/10/21 & 52568.811--52568.832 & 1.5 & 0.5 & FHXR & 0.215 (Ry) \\
77 & 70062-01-03-00 & 02/10/21 & 52568.879--52568.897 & 1.4 & 0.4 & FHXR & 0.190 (Ry) \\
78 & 70062-01-02-03 & 02/10/21 & 52568.948--52568.963 & 1.1 & 0.4 & FHXR & 0.202 (Ry) \\
79 & 70062-01-03-02 & 02/10/21 & 52568.985--52569.029 & 2.0 & 0.7 & FHXR & 0.101 (Ry) \\
80 & 70062-03-01-00 & 02/12/22 & 52630.102--52630.197 & 5.7 & 1.9 & FHXR & 0.073 (Ry) \\
81 & 70062-03-02-00 & 02/12/23 & 52631.222--52631.327 & 6.2 & 2.2 & FHXR & 0.156 (Ry) \\
\hline
\end{tabular}
\end{center}
\label{list2}
\end{table*}%

\begin{table*}
\contcaption{}\begin{center}
\begin{tabular}{|lccccccc}
\hline
\hline
No. & ObsID & Date & MJD & PCA & HEXTE & X-ray/ & Radio \\
& & & Interval & Exp. & Exp. & Radio & Flux \\
& & [yy/mm/dd] & [d] & [ks] & [ks] & State & [Jy] \\
\hline
\hline
82 & 91412-02-01-00 & 05/05/10 & 53500.605--53500.629 & 2.0 & 0.6 & T & 0.080 (Ry) \\
\hline
83 & 91090-04-01-00 & 06/01/25 & 53760.430--53760.467 & 3.0 & 1.0 & HYS & 0.004 (Ry) \\
84 & 91090-04-02-00 & 06/01/25 & 53760.493--53760.530 & 3.2 & 1.1 & HYS & 0.007 (Ry) \\
85 & 91090-04-03-00 & 06/01/25 & 53760.560--53760.597 & 3.1 & 1.0 & HYS & 0.007 (Ry) \\
86 & 91090-05-01-00 & 06/01/26 & 53761.083--53761.269 & 9.6 & 3.3 & HYS & 0.044 (Ry) \\
87 & 91090-05-02-00 & 06/01/26 & 53761.412--53761.449 & 3.1 & 1.1 & HYS & 0.028 (Ry) \\
88 & 91090-05-03-00 & 06/01/26 & 53761.671--53761.711 & 3.2 & 1.0 & HYS & 0.016 (Ry) \\
89 & 91090-05-04-00 & 06/01/26 & 53761.737--53761.771 & 2.4 & 0.7 & HYS & 0.015 (Ry) \\
90 & 91090-05-05-00 & 06/01/27 & 53762.064--53762.248 & 9.6 & 3.3 & HYS & 0.004 (Ry) \\
91 & 91090-06-01-00 & 06/01/28 & 53763.047--53763.232 & 9.7 & 3.3 & HYS & 0.003 (Ry) \\
92 & 91090-06-02-00 & 06/03/11 & 53805.500--53805.537 & 3.1 & 1.0 & FIM & 2.961 (Ry) \\
93 & 91090-06-02-01 & 06/03/11 & 53805.630--53805.668 & 3.2 & 1.0 & FIM & 1.766 (Ry) \\
94 & 91090-06-03-00 & 06/03/11 & 53805.826--53805.946 & 6.3 & 2.1& FIM & 1.783 (Ry) \\
95 & 91090-02-01-00 & 06/05/12 & 53867.006--53867.040 & 2.8 & 1.0 & FHXR & 7.498 (Ry) \\
96 & 91090-02-01-06 & 06/05/12 & 53867.725--53867.777 & 3.2 & 1.1 & FIM & 6.348 (Ry) \\ 
97 & 91090-02-01-01 & 06/05/13 & 53868.052--53868.093 & 3.2 & 1.1 & FHXR & 6.558 (Ry) \\
98 & 91090-02-01-07 & 06/05/13 & 53868.181--53868.271 & 5.1 & 1.5 & FHXR & 5.519 (Ry) \\
99 & 91090-02-01-02 & 06/05/13 & 53868.315--53868.354 & 3.2 & 0.9 & FHXR & 5.160 (Ry) \\
100 & 91090-02-01-03 & 06/05/13 & 53868.354--53869.214 & 3.2 & 1.0 & FHXR & 6.678 (Ry) \\
101 & 91090-02-01-05 & 06/05/13 & 53868.836--53868.894 & 3.2 & 1.1 & FHXR & 7.174 (Ry) \\ 
102 & 91090-02-01-08 & 06/05/14 & 53869.558--53869.599 & 2.3 & 0.8 & FIM & 6.316 (Ry) \\
103 & 91090-02-01-04 & 06/05/14 & 53869.820--53869.871 & 3.2 & 1.1 & FIM--FHXR & 5.745 (Ry) \\ 
104 & 91090-02-01-10 & 06/05/15 & 53870.017--53870.057 & 3.2 & 1.1 & FIM & 5.831 (Ry) \\
105 & 91090-02-01-12 & 06/05/15 & 53870.214--53870.254 & 3.2 & 0.9 & FIM & 4.829 (Ry) \\
106 & 91090-02-01-11 & 06/05/15 & 53870.603--53870.649 & 2.6 & 0.9 & FIM & 3.643 (Ry) \\
107 & 91090-02-01-09 & 06/05/15 & 53870.998--53871.039 & 3.3 & 1.1 & FIM & 1.869 (Ry) \\
108 & 91090-03-01-00 & 06/05/17 & 53872.052--53872.218 & 9.4 & 2.8 & FIM & 1.279 (Ry) \\
109 & 91090-03-02-00 & 06/05/17 & 53872.766--53872.947 & 9.8 & 3.3 & FIM & 1.120 (Ry) \\
110 & 91090-01-01-000 & 06/07/26 & 53942.458--53942.699 & 13.9 & 4.1 & FHXR & 2.797 (RA) \\
111 & 91090-01-01-00 & 06/07/26 & 53942.727--53942.896 & 7.4 & 2.5 & FHXR & 2.797 (RA) \\
112 & 91090-01-01-01 & 06/07/26 & 53942.937--53942.959 & 1.8 & 0.6 & FHXR & 2.797 (RA) \\
\hline
113 & 93434-01-01-00 & 08/04/23 & 54579.220--54579.290 & 3.6 & 1.1 & FIM & 0.227 (RA) \\
114 & 93434-01-02-00 & 08/04/25 & 54581.098--54581.120 & 1.9 & 0.6 & FSXR & 0.183 (RA) \\
115 & 93434-01-02-01 & 08/04/25 & 54581.228--54581.251 & 1.9 & 0.5 & FSXR & 0.183 (RA) \\
116 & 93434-01-03-00 & 08/04/27 & 54583.059--54583.082 & 1.8 & 0.6 & FIM & 0.485 (RA) \\
117 & 93434-01-03-01 & 08/04/27 & 54583.212--54583.232 & 1.6 & 0.4 & FSXR & 0.485 (RA) \\
\hline
118 & 94328-01-01-01 & 09/01/10 & 54841.899--54841.925 & 2.2 & 0.7 & T & 0.205 (Ry) \\
119 & 94328-01-01-00 & 09/01/10 & 54841.995--54842.065 & 6.0 & 2.1 & T & 0.205 (Ry) \\
120 & 94328-01-02-00 & 09/01/25 & 54856.169--54856.327 & 7.5 & 2.5 & T & -- \\
121 & 94328-01-03-00 & 09/02/08 & 54870.948--54871.097 & 7.8 & 2.6 & T--FHXR & -- \\
122 & 94328-01-04-00 & 09/02/23 & 54885.869--54886.017 & 7.7 & 2.6 & T & -- \\
123 & 94328-01-05-00 & 09/03/09 & 54899.081--54899.123 & 3.3 & 1.0 & T & -- \\
124 & 94328-01-05-01 & 09/03/09 & 54899.144--54899.189 & 3.5 & 1.0 & T & -- \\
125 & 94328-01-06-00 & 09/03/21 & 54911.969--54912.056 & 4.7 & 1.6 & T & -- \\
126 & 94328-01-06-01 & 09/03/22 & 54912.174--54912.204 & 2.4 & 0.8 & T & -- \\
127 & 94328-01-07-00 & 09/04/04 & 54925.899--54926.040 & 7.2 & 2.3 & T & -- \\
128 & 94328-01-08-00 & 09/04/20 & 54941.084--54941.113 & 2.2 & 0.6 & FHXR-T & -- \\
129 & 94328-01-08-02 & 09/04/20 & 54941.135--54941.179 & 3.4 & 1.0 & FHXR-T & -- \\
130 & 94328-01-08-01 & 09/04/20 & 54941.267--54941.280 & 1.1 & 0.2 & FHXR-T & -- \\
131 & 94328-01-09-00 & 09/05/04 & 54955.862--54956.016 & 7.3 & 2.3 & FHXR & -- \\
132 & 94328-01-10-00 & 09/05/16 & 54967.727--54967.871 & 7.3 & 2.4 & FHXR & -- \\
133 & 94328-01-11-00 & 09/05/31 & 54982.739--54982.883 & 7.4 & 2.4 & FIM & -- \\
134 & 94328-01-12-01 & 09/06/14 & 54996.542--54996.604 & 2.8 & 0.9 & FIM & -- \\
135 & 94328-01-12-00 & 09/06/14 & 54996.777--54996.850 & 4.0 & 1.2 & FIM & -- \\
\hline
\end{tabular}
\end{center}
\label{list3}
\end{table*}%

\label{lastpage}

\end{document}